\documentclass[]{elsarticle}
\usepackage{graphicx}
\usepackage{amssymb}
\usepackage{amsthm}
\usepackage{lineno}
\usepackage{mathrsfs,amsfonts,amsmath,algorithm}
\usepackage{subcaption}
\usepackage{array}
\usepackage{xcolor}
\usepackage{hyperref}
\usepackage{ulem}
\usepackage{makecell}
\usepackage{bm}% bold math
\usepackage{threeparttable}

\journal{Journal of Computational Physics}

\newcommand{\dd}[2]{\frac{d #1}{d #2}}
\newcommand{\pp}[2]{\frac{\partial #1}{\partial #2}}
\newcommand{\ip}[1]{\left\langle #1 \right\rangle} %for inner product
\newcommand{\avg}[1]{ #1 _{avg}} %for inner product
\newcommand{\R}{\mathbb{R}}
\newcommand{\mat}[1]{\begin{bmatrix}#1\end{bmatrix}}
\newcommand{\mus}{{m_{us}}}
\newcommand{\integrate}{\int_{0}^{T}}

\newcommand{\az}{\overline{\zeta}}
\newcommand{\aZ}{\overline{Z}}

\newcommand{\av}{\overline{v}}
\newcommand{\aw}{\overline{w}}
\newcommand{\aW}{\overline{W}}

\newcommand{\as}{\overline{\mathcal{S}}}

\newcommand{\intT}{\int_0^T}
\newcommand{\avgT}{\lim_{T\to\infty}\frac{1}{T} \intT}

\theoremstyle{definition}

\newtheorem{defn}{Definition}
\newtheorem{thm}{Theorem}

\newcommand\reout{\bgroup\markoverwith{\textcolor{red}{\rule[0.5ex]{2pt}{0.4pt}}}\ULon}
\newcommand\reeout{\bgroup\markoverwith{\textcolor{blue}{\rule[0.5ex]{2pt}{0.4pt}}}\ULon}
\newcommand\naxout{\bgroup\markoverwith{\textcolor{green}{\rule[0.5ex]{2pt}{0.4pt}}}\ULon}

\begin{document}

\begin{frontmatter}

\title{Adjoint sensitivity analysis on chaotic dynamical systems by Non-Intrusive Least Squares Adjoint Shadowing (NILSAS)}

\author[Angxiu]{Angxiu Ni}\corref{cor}
\cortext[cor]{Corresponding author.}
\ead{niangxiu@gmail.com}
\ead[url]{https://math.berkeley.edu/~niangxiu/}
\address[Angxiu]{Department of Mathematics, University of California, Berkeley, CA 94720, USA}

\author[Chai]{Chaitanya Talnikar}
\ead{chaitukca@gmail.com}
\address[Chai]{Nvidia Corporation, Santa Clara, CA 95051, USA}

\begin{abstract}
We develop the NILSAS algorithm, 
which performs adjoint sensitivity analysis of chaotic systems via computing the adjoint shadowing direction.
NILSAS constrains its minimization to the adjoint unstable subspace, and can be implemented with little modification to existing adjoint solvers.
The computational cost of NILSAS is independent of the number of parameters.
We demonstrate NILSAS on the Lorenz 63 system and a weakly turbulent three-dimensional flow over a cylinder.
\end{abstract}


\end{frontmatter}
%\linenumbers

\section{Introduction}

Sensitivity analysis is a powerful tool in helping scientists and engineers design products \cite{Jameson1988,Reuther},
control processes and systems \cite{Bewley2001,Bewley2001a}, solve inverse problems \cite{Tromp}, 
estimate simulation errors \cite{Becker2001,Giles2002,Fidkowski,Persson2009_mesh_adapt}, 
assimilate measurement data \cite{Thepaut1991,COURTIER1993} 
and quantify uncertainties \cite{Marzouk2015}.

Conventional sensitivity analysis works well for systems with a stable fixed point or a periodic orbit.
However, when the system is chaotic and the design objective is a long-time-averaged quantity, 
conventional sensitivity methods, be it tangent or adjoint, fails to provide useful sensitivity information.
Chaotic dynamical systems are typically modeled as hyperbolic systems,
for which many sensitivity analysis methods have been developed, such as by Lea et al. in \cite{Lea2000,eyink2004ruelle},
by Abramov and Majda in \cite{abramov2007blended,Abramov2008},
and by Lucarini et al. in \cite{lucarini_linear_response_climate,lucarini_linear_response_climate2}.

Sensitivity of hyperbolic dynamical systems can be computed via the shadowing direction.
Bowen \cite{Bowen_shadowing} proved that for a trajectory of a uniform hyperbolic systems, if we introduce a small perturbation in the governing equation,
there exists a shadowing trajectory, which satisfies the perturbed equations, but still lies close to the base trajectory.
Later Pilyugin \cite{Pilyugin_shadow_linear_formula} gave a formula for the first order difference 
between the shadowing trajectory and the base trajectory.
In this paper we call such first order difference the shadowing direction.
The Least Squares Shadowing (LSS) approach, developed by Wang et al., \cite{wang2014convergence,Wang_ODE_LSS,Chater_convergence_LSS}, 
computes such shadowing direction through an $L^2$ minimization, and use it for sensitivity analysis.

The Non-Intrusive Least Squares Shadowing method (NILSS) developed by Ni et al. \cite{Ni_NILSS_AIAA_2016,Ni_NILSS_JCP}
finds a `non-intrusive' formulation of the LSS problem which allows constraining the minimization problem in LSS to the unstable subspace.
For many real-life problems, the dimension of unstable subspace is much lower than the dimension of the dynamical system, 
and NILSS can be thousands times faster than LSS. 
A major variant of NILSS is the Finite Difference NILSS (FD-NILSS) algorithm \cite{Ni_fdNILSS},
whose implementation requires only primal solvers, but not tangent solvers.
FD-NILSS has been applied to several complicated flow problems 
\cite{Ni_CLV_cylinder,Ni_fdNILSS} which were too expensive for previous sensitivity analysis methods.

The marginal cost for a new parameter in NILSS is only computing one extra inhomogeneous tangent solution.
Yet for cases where there are many parameters and only a few design objectives, an adjoint version of NILSS is desired.
A continuous adjoint version appeared in the first publication of NILSS \cite{Ni_NILSS_AIAA_2016}; 
however, this version of lacks the constraint on the neutral subspace, which will be explained in our current paper.
Blonigan \cite{Blonigan_2017_adjoint_NILSS} developed a discrete adjoint version of NILSS, 
which was later implemented by Chandramoorthy et al. \cite{Chandramoorthy2018_nilss_ad} using automatic differentiation.
In comparison to this discrete adjoint NILSS, NILSAS does not require tangent solvers, and requires less modification to existing adjoint solvers,
and the simplicity of the formula of NILSAS should also give it more robustness and perhaps better convergence.

Recently, we defined the adjoint shadowing direction for both hyperbolic flows and diffeomorphisms,
which is a bounded inhomogeneous adjoint solution with several other properties \cite{Ni_adjoint_shadowing}.
We showed that the adjoint shadowing direction exists uniquely on a given trajectory, and can be used for adjoint sensitivity analysis.
Adjoint shadowing direction is defined using only adjoint flows, giving us a chance to get rid of tangent solvers in our algorithm, and arrive at a neat formula.

This paper presents the Non-Intrusive Least Squares Adjoint Shadowing (NILSAS) algorithm, 
where we construct a least squares problem to approximate the adjoint shadowing direction and then compute adjoint sensitivity.
The `non-intrusive' formulation of NILSAS allows it be built using existing adjoint solvers,
and more importantly, it allows the minimization be constrained to the unstable adjoint subspace.
The main body of this paper will be about continuous dynamical systems, 
and we briefly discuss NILSAS for discrete systems in \ref{s:discrete systems}.

We organize the rest of this paper as follows.
First, we prepare our study by defining our problem and reviewing adjoint flows and adjoint shadowing directions; we also provide some intuitions to help understanding adjoint shadowing directions.
Then we derive the NILSAS algorithm.
Then we present a detailed procedure list for our algorithm and give several remarks on the algorithm.
Finally, we demonstrate NILSAS on the Lorenz 63 system and a weakly turbulent three-dimensional (3D) flow over a cylinder.

\section{Preparations}

The governing equation of a hyperbolic flow, which models a time-evolving chaotic process, is: 
\begin{equation} \label{e:primal_system}
  \dd{u}{t} = f(u,s), \quad u(t=0) = u_0\;.
\end{equation}
This differential equation is called the primal system, and a solution $u(t)$ is the primal solution.
Here $f(u,s):\R^m\times \R\rightarrow\R^m$ is a smooth function, $u\in \R^m$ is the state of the dynamical system, $u_0$ is the initial condition, and $s\in\R$ is the parameter.
For now we assume there is only one parameter;
and in section~\ref{s:remarks misc} we will explain how we can compute sensitivities with respect to several parameters with almost no additional cost.

In this paper, we assume there is only one objective, which is a long-time-averaged quantity.
To define it, we first let $J(u,s):\R^m\times\R\rightarrow\R$ be a continuous function that represents the instantaneous objective.
The objective is obtained by averaging over a semi-infinite trajectory:
\begin{equation} \label{eq:average J}
  \avg{J}:= \lim\limits_{T\rightarrow\infty} \frac{1}{T}\integrate J(u,s)dt \,.
\end{equation}
We assume the system has a global attractor \cite{Chater_convergence_LSS}, hence $ \avg{J} $ only depends on $s$.
Our goal of this paper is to develop an algorithm computing the sensitivity, $d\avg{J} / d s$, whose marginal cost for a new parameter is negligible.

\subsection{Adjoint flow}

\begin{defn} \label{d:adjoint flow operator}
  A homogeneous adjoint solution $w(t):\R \rightarrow \R^m$ is a function which solves the homogeneous adjoint equation:
  \begin{equation} \label{e:homogeneous adjoint}
    \dd wt + f_u ^T w =0,
  \end{equation}
  where $\cdot^T$ is the matrix transpose.
  An inhomogeneous adjoint solution is a function $v(t):\R \rightarrow \R^m$ which solves:
  \begin{equation}\label{e:inhomogeneous adjoint}
    \dd {v}t + f_u ^T v =g(t),
  \end{equation}
  where $g(t):\R \rightarrow \R^m$ is a vector-valued function of time.
\end{defn}

In numerical implementations, we typically solve adjoint equations backward in time.
This is because, as shown in \cite{Ni_adjoint_shadowing}, when solving backward in time, 
the dimension of the unstable adjoint subspace is the same as the unstable tangent subspace, which is typically much lower than $m$.
On the other hand, if we solve the adjoint equation forward in time,
the unstable subspace has much higher dimension, causing strong numerical instability.

\begin{defn}
  In this paper, an adjoint covariant Lyapunov vector (CLV) with adjoint Lyapunov exponent (LE) $\lambda$
  is a homogeneous adjoint solution $\az(t)$ such that there is a constant $C$, for any $t_1,t_2\in \R$, 
  \begin{equation}
    \| \az (t_1)\|\le C e^{\lambda(t_2-t_1)} \|\az(t_2)\| \;.
  \end{equation}
\end{defn}

%  time direction is reversed, how to sort CLV, history knowledge
Note that the time direction in the above definition is reversed: 
if the adjoint CLV grows exponentially backward in time, its exponent is positive.
Adjoint CLVs with positive exponents are called unstable, those with negative exponents are stable, 
and those with zero exponent is are neutral.
In this paper, we sort adjoint CLVs by descending order of their exponents.
  The earliest mention of adjoint CLVs was by Kuptsov and Parlitz in \cite{Kuptsov2012_define_adjoint_CLV}.
  For the purpose of defining adjoint shadowing directions and deriving the NILSAS method, 
  we recently proved the existence of adjoint CLVs under the same assumptions of adjoint shadowing theorem, 
  and found some relation between CLVs and adjoint CLVs.

% relation to tangent CLVs
  Adjoint CLVs are homogeneous adjoint solutions whose norm grows exponentially, and the adjoint LE is measured backward in time.
  This is similar but also different from tangent CLVs, which are tangent solutions measured forward in time.
  The CLV structure for the adjoint flow is the same as the tangent flow.
  That is, the adjoint LE spectrum is the same as the tangent LE spectrum.
  Moreover, the subspace of CLVs with an exponent $\lambda$ is perpendicular to the subspace of all adjoint CLVs with exponents not $\lambda$, and vice versa.
  If we can write the full set of CLVs as a matrix valued function of time, $W(t)$, then $W^{-T}(t)$, where $\cdot^{-T}$ is the inverse of transpose, is a matrix whose columns are adjoint CLVs:
  readers can verify that $W^{-T}(t)$ satisfies the properties listed above.
  Note that we do not know if CLVs and adjoint CLVs with the same exponent are perpendicular or parallel.

% adjoint CLV structure: relation to tangent CLV, uniform hyperbolicity 
We assume our system is uniform hyperbolic and it has a bounded global attractor.
Definition of hyperbolicity can be found in most textbook on dynamical system such as \cite{Katok_thick_book},
and readers may also refer to \cite{Ni_adjoint_shadowing} for a definition using the same notation as this paper.
Uniform hyperbolicity requires that the tangent space can be split into stable subspace, unstable subspace, and a neutral subspace of dimension one.
Together with the boundedness of the attractor, 
we can show the angles between two subspaces of different sets of tangent CLVs are always larger than some positive angle.
Since the adjoint equations have the same structure as the tangent ones,
there is only one neutral adjoint CLV, and adjoint CLVs are always bounded away from each other \cite{Ni_adjoint_shadowing}.

\subsection{Adjoint shadowing directions}\label{s:adjoint_shadowing_direction}

In \cite{Ni_adjoint_shadowing}, the author defined adjoint shadowing directions, 
proved their unique existence on a given trajectory, 
and showed how to use them for adjoint sensitivity analysis.
We briefly restate the main results in this subsection.

% main definition
\begin{defn}
  On a trajectory $u(t)$ on the attractor, for $ t\ge 0$,
  the adjoint shadowing direction $\av^\infty: \R_+\rightarrow\R^m$ is defined as a function with the following properties:
  \begin{enumerate}
    \item $\av^\infty$ solves the inhomogeneous adjoint equation:
      \begin{equation} \label{e:av solve inhomo}
        \dd {\av^\infty}t + f_u^T \av^\infty = - J_u \,,
      \end{equation}
      where subscripts are partial derivatives, that is, $f_u = \partial f /\partial u$, $J_u = \partial J /\partial u$.
    \item $\av^\infty(t=0)$ has zero component in the unstable adjoint subspace. 
    \item $\|\av^\infty(t)\|$ is bounded by a constant for all $t\in\R_+$.
    \item The averaged inner-product of $\av^\infty$ and $f$ is zero:
      \begin{equation} \label{e:av_f_zero}
        \ip{\av^\infty, f}_{avg} := \lim_{T\rightarrow \infty} \frac 1T \int_0^T \ip{\av^\infty(t), f(t)} = 0 \,,
      \end{equation} 
      where $\ip{\cdot,\cdot}$ is the inner-product on the Euclidean space.
  \end{enumerate}
\end{defn}

  We remind readers to distinguish the three different kinds of adjoint solutions we mentioned: homogeneous adjoint solutions, inhomogeneous adjoint solutions and adjoint shadowing directions.
  Homogeneous adjoint solutions are different from inhomogeneous ones, since homogeneous adjoint equations must have zero right-hand-sides.
  The adjoint shadowing direction is an inhomogeneous adjoint solution, but not any inhomogeneous adjoint solution:
  it must in extra have three more properties listed in the definition.
  In fact, one way to view NILSAS is that we search the space of all inhomogeneous adjoint solutions to find one such that it mimics the other three properties.
  More specifically, we minimize the $L^2$ norm, and constrain the inner product with $f$: this derivation will be revealed in later sections.

% main theorem
\begin{thm}\label{thm:flow}
  For a uniform hyperbolic system with a global compact attractor,
  on a trajectory on the attractor,
  there exists a unique adjoint shadowing direction.
  Further, we have the adjoint sensitivity formula:
  \begin{equation}\label{e:adjoint sensitivity}
    \dd {\avg{J}} {s} = \lim_{T\rightarrow\infty} \frac 1T \integrate \ip{\av^\infty, f_s} + J_s \, dt \,.
  \end{equation}
\end{thm}

  We explain the assumption of the adjoint shadowing theorem.
  First, if a dynamical system has a compact global attractor, it means there is a bounded set of states, or the attractor, 
  such that no matter what initial condition the system starts from, the trajectory will eventually enter the attractor and never leave.
  Second, uniform hyperbolicity here mainly means that there is only one neutral CLV.
  Third, by the compactness, the angles between all CLVs are larger than a positive angle, regardless of where we are on the attractor.

  Why do we make above assumptions in theories for shadowing methods?
  The main reason for assuming only one neutral CLV in shadowing methods is to prevent linear growth in inhomogeneous tangent/adjoint solutions.
  The main reason for global attractability is to ensure that shadowing trajectories are representative of the averaged behavior of the system.
  The main reason for compactness is because we want a bound for the projection operators projecting onto a particular subspace.
  Still, we remind readers that, in practice, shadowing methods may be effective beyond above assumptions, 
  as to be discussed in section~\ref{s:remarks misc}.

Rather than giving an explicit expression of adjoint shadowing directions, which can be found as well in \cite{Ni_adjoint_shadowing},
the definition is stated as a criterion,
where we check several properties to determine if a function is indeed the adjoint shadowing direction.
In fact, we forged this definition for designing the NILSAS algorithm, which will be revealed in the next section.

\subsection{Interpreting adjoint shadowing directions}

We give several different perspectives to help readers build intuitions on adjoint shadowing directions.
To start with, we revisit some formal descriptions of shadowing operators.
The shadowing operator, denoted as $\mathcal{S}$, 
can be viewed roughly as mapping a vector-valued function $f_s(t)$ to another vector-valued function $v^\infty(t)$,
Here $f_s(t)$ is the perturbation on $f$ due to parameter perturbations;
$v^\infty$ is the (tangent) shadowing direction, 
which is first order approximation of the difference between the shadowing and the base trajectory.
Note that $\mathcal{S}$ is a linear operator, and both $f_s(t)$ and $v^\infty(t)$ are linear approximations.
If we neglect the subtleties due to the neutral CLV, we have roughly
\begin{equation}
  \dd {\avg{J}} {s} \approx \ip{v^\infty, J_u}_{avg}  =  \ip{\mathcal{S}(f_s), J_u}_{avg} .    
\end{equation}
Where $\ip{\cdot, \cdot}_{avg}$ is an inner product.

First we provide a utility point of view for adjoint shadowing directions, which is also an algebraic point of view.
Riesz's representation theorem tells us that there is an adjoint operator $\as$ such that 
\begin{equation}
  \ip{\mathcal{S}(f_s), J_u}_{avg} = \ip{f_s, \as(J_u)}_{avg} . 
\end{equation}
The adjoint shadowing direction, $\av^\infty$, can be viewed as $\as(J_u)$.
Suppose now that we have a computer program which approximately functions as $\as$.
If we have two parameters $s_1, s_2$, then they can perturb the governing equation by $f_{s_1}$ and $f_{s_2}$, whereas $J_u$ keeps the same.
This means that we only need to run our adjoint program once, and use the result to inner-product with both $f_{s_1}$ and $f_{s_2}$.
For cases where there are many parameters $s$ and a few objectives $J$, we only need to run our program a few times.

Then a physical point of view.
Formally $dJ_{avg}/ds = \ip{\partial f /\partial s, \av^\infty} $,
and we can formally cancel $ds$ on both side and divide by $df$.
Thus we get $\av^\infty = \partial J_{avg}/\partial f$.
This means that the adjoint shadowing direction can be viewed as how perturbations in the output of $f$ affects the objective.
For example, in our fluid examples, if the $\rho E$ component of $\av^\infty$ has value $\beta$ for all $u$, 
it means that if we can somehow keep infusing unit energy into the flow, the objective will be changed by $\beta$.
Note that in $\av^\infty$ we do not prescribe how perturbations in $f$ is generated from changing a particular parameter; the dependence $f$ on $s$ is described in another term, $f_s$.

Then an implementation point of view.
The adjoint operator of a matrix is simply its transpose.
For many other cases, we can see that matrix transposition often appears as a crucial step in the formulation of adjoint operators.
So once we derive an new adjoint formula of something, we may ask if it can be presented as neatly as transposing a matrix.
Adjoint shadowing directions satisfy inhomogeneous adjoint equations, which is a linear ODE whose matrix is the transpose of the Jacobian matrix.
Hence there is chance that algorithms, such as NILSAS, do not differ too much from existing adjoint solvers.
In fact, giving a neat recipe for adjoint shadowing direction is the main contribution of both the adjoint shadowing theorem and the NILSAS algorithm,
since the existence is already given by Riesz's representation theorem.

Above different views on adjoint shadowing directions are the best intuitions we can come up with now.
Still, we acknowledge that it is not yet as intuitive as shadowing direction,
which can be viewed as vectors `starting from' a base trajectory, and `pointing to' a shadowing trajectory, 
with parameter $s+\Delta s$, that truly exists.
Moreover, our intuition is not quite refined yet, as we still need to draw upon the abstract notion of duality with their tangent counterparts,
in order to describe the subtle difference among all those adjoint solutions: 
homogeneous adjoint solutions, inhomogeneous adjoint solutions and adjoint shadowing directions.
We humbly ask readers to inform us better intuitions.

\section{Deriving NILSAS}\label{s:derive_NILSAS}

\subsection{The non-intrusive formulation}

% summarize
On a finite trajectory of time span $[0,T]$, the NILSAS algorithm computes a $\av$ which approximates $\av^\infty$.
Since the definition of the adjoint shadowing direction is similar to the tangent shadowing direction,
it is not surprising that NILSAS has similar formulation as NILSS.

In NILSAS, we strictly enforce the first property of adjoint shadowing directions by constraining our solutions to inhomogeneous adjoint solutions.
The second property is changed to a symmetric statement that stable component in $\av(T)$ should be $O(1)$, 
which can be easily satisfied.
The third property is approximated by minimizing the $L^2$ norm of the inhomogeneous adjoint solution.
The fourth property is strictly enforced by adding a constraint to our minimization problem.
In this subsection, we explain why the $\av$ given by this reverse-engineering is a good approximation of $\av^\infty$.

% first property -> representation of av 
Our algorithm strictly enforces the first property of $\av^\infty$.
To do this, we represent the solution set of equation~\eqref{e:av solve inhomo} as a particular solution plus the space of homogeneous solutions.
We select the particular solution as the conventional inhomogeneous adjoint solution $\av^*$, which is defined as the solution of:
\begin{equation}\label{e:conventional adjoint}
  \dd {\av^*}t + f_u ^T \av^* =-J_u\,, \; \av^*(T)=0\;.
\end{equation}
Then we select the collection of all adjoint CLVs, $\aZ = [\az_1,\cdots,\az_m]$, 
all of which have terminal condition $\|\az_j(T)\|=1$, as the basis of the space of homogeneous solutions.
Hence we can enforce the first property by considering candidates only in the following form for some $\left\{a_j\right\}_{j=1}^m$:
\begin{equation}\label{e:av_as_v*_plus_CLV}
  \av = \av^* + \sum_{j=1}^m a_j \az_j = \av^* + \aZ a \;,
\end{equation}
where the coefficients $a = [a_1,...,a_m]^T$ is a column vector.
Another interpretation of our way of enforcing the first property is that, 
we want to start $\av$ from $\av^*$, and modify by adding adjoint CLVs to approximate $\av^\infty$;
in other words, the coefficients $a$ should be such that $Za \approx \av^\infty - \av^*$.
We then use other properties to determine the coefficients for stable, unstable and neutral CLVs.

% property 2 -> coefficients for stable CLVs
When defining adjoint shadowing directions in \cite{Ni_adjoint_shadowing}, the author was considering functions defined starting from time zero,
whereas in our case here, adjoint solutions are solved from $T$ backward in time.
Hence, in order to keep the sensitivity formula, we change the second property to a symmetric statement,
that is, we want the stable component in $\av(T)$ be in the order $O(1)$, meaning be bounded by a constant independent of $T$.
Equivalently, we require coefficients for stable CLVs be $O(1)$.
Another way to interpret is that, if this $O(1)$ condition is true, 
then since $\av^\infty(T)$ is $O(1)$, the stable component in $\av(T)-\av^\infty(T)$ is $O(1)$; 
now since stable CLVs decay exponentially fast, their contribution in $\av - \av^\infty$ can be neglected.
This $O(1)$ condition is a loose requirement and, as we will see, it can be easily satisfied.

% property 3 -> minimization over only unstable directions
To mimic the boundedness in the third property of adjoint shadowing directions, we minimize $\|\av\|_{L^2}$,
which determines the coefficients for the unstable CLVs. 
Indeed, this minimization removes significant unstable CLVs from $\av-\av^\infty$, 
since otherwise this difference would grow exponentially, 
and since $\av^\infty$ is bounded, $\av$ would have large $L^2$ norm.

% property 4 -> determine the coefficient for neutral adjoint CLV
We strictly enforce the fourth property of $\av$.
This determines the coefficient for the neutral adjoint CLV, 
since as shown in \cite{Ni_adjoint_shadowing},
$f$ is always orthogonal to non-neutral adjoint CLVs.
Note also that the norm of the neutral adjoint CLV is bounded, unlike neutral tangent CLV, which can have linear growth.
Hence we can allow its coefficient be $O(1)$ without jeopardizing the boundedness property we used earlier.
In fact, the adjoint NILSS in \cite{Ni_NILSS_AIAA_2016} lacks exactly this constraint on the neutral adjoint CLV.

% approximate unstable CLVs by adjoint solutions
To summarize, we determine coefficients for unstable adjoint CLVs via a minimization,
the coefficient for the neutral adjoint CLV via equation~\eqref{e:av_f_zero},
and we do not care coefficients for stable adjoint CLVs too much.
Hence there is no need to provide stable CLVs to our algorithm;
it is even unnecessary to provide accurate non-stable CLVs, they can contain some stable components at $T$.
Further, we care not individual CLVs but only their span.
Hence we can replace $\aZ$ by $\aW=[\aw_1,\cdots,\aw_M]$, 
with $M\ge\mus+1$, $\mus$ being the number of unstable CLVs,
and $\{\aw_j\}_{j=1}^M$ are homogeneous adjoint solutions whose non-stable components at $T$ span the entire non-stable subspace.
Such a set of solutions can be obtained by solving homogeneous adjoint equations from almost all terminal conditions of $M$ randomized unit vectors.

% summary: give the formula, remark non-intrusiveness
With above discussions, we see that our algorithm should solve the NILSAS problem on one segment:
\begin{equation} \label{e:NILSAS_1_segment} \begin{split}
  \min_{a\in \R^M} &\;\frac{1}{2} \integrate \ip{\av^* + \aW a, \av^* + \aW a} \;,\\
                   &\textnormal{s.t. } \int_0^T \ip{\av^* + \aW a, f} = 0 \;.
\end{split}\end{equation}
This is simply a least squares problem with arguments $a\in \R^M$.
Note that all adjoint solutions can be computed after making little modifications to existing adjoint solvers, 
and our minimization is constrained to essentially only the unstable adjoint subspace:
these are the benefits of the non-intrusive formulation. 
Letting $\av = \av+\aW a$, we can compute sensitivity via equation~\eqref{e:adjoint sensitivity} on a finite trajectory:
\begin{equation}
  \dd {\avg{J}} {s} \approx \frac 1T \integrate \ip{\av, f_s} + J_s \, dt \,.
\end{equation}

\subsection{Dividing trajectory into segments}\label{s:divide trajectory}

% describe the issue 
An issue in numerical stability is that, as the trajectory gets longer, 
all adjoint solutions become dominated by the fastest growing adjoint CLV; 
as a result, the minimization problem in equation~\eqref{e:NILSAS_1_segment} becomes ill-conditioned.
This issue also happened in NILSS \cite{Ni_NILSS_JCP,Ni_NILSS_AIAA_2016} and in the algorithm for computing CLVs \cite{Ginelli2013_CLV},
and here we use a similar technique to resolve it, that is, dividing the whole trajectory into multiple segments, and rescaling at interfaces.

% vague description of how dividing into segments solves the problem.
Roughly speaking, at the end of each segment,
we orthogonalize and rescale adjoint solutions $\aW$ and $\av^*$ so that they are no longer dominated by the first CLV.
Note here since adjoint solutions are integrated backward in time, the initial condition is at the end of a segment.
Despite that now $\aW$ and $\av^*$ are discontinuous across segments, we can still construct $\av$ as their linear combinations,
and keep $\av$ continuous across all segments.
This continuous $\av$ computed from multiple segments should be identical to that solved on one large segment containing the entire trajectory.

\begin{figure}[ht]
  \centering
  \includegraphics[width=0.6\textwidth]{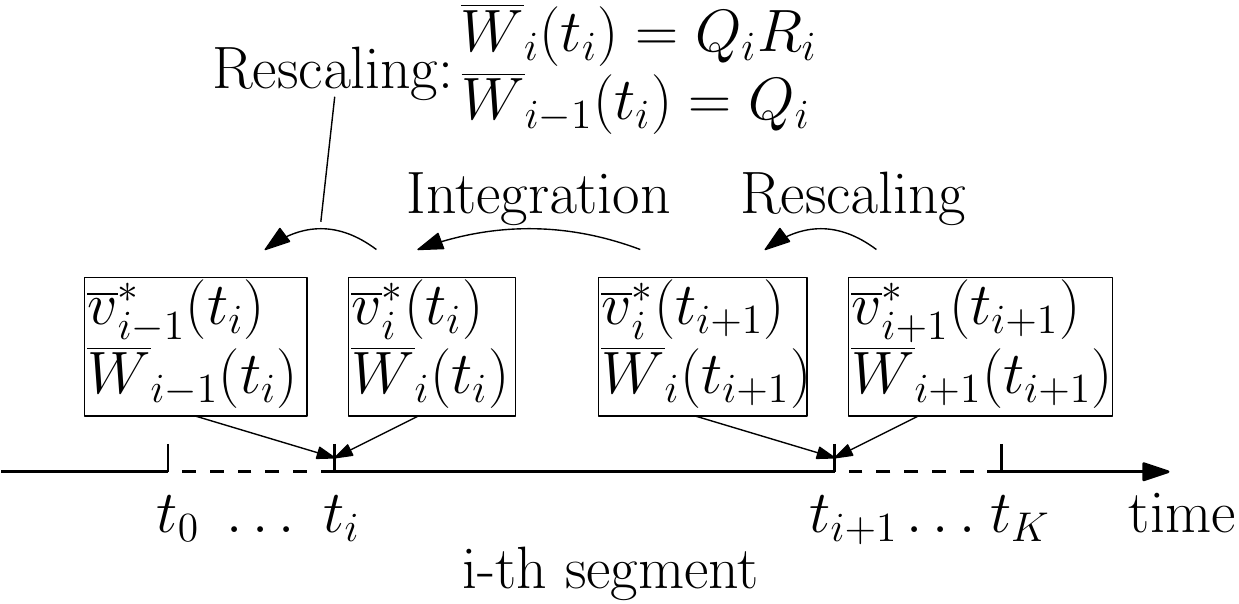}
  \caption{Notations for multiple segments.
  $\aW_i(t), \av^*_i(t)$ are defined on the $i$-th segment, which spans $t\in[t_i, t_{i+1}]$; $Q_i, R_i$ are defined at $t_i$.
  `Integration' refers to integrating adjoint equations for $\aW_i(t), \av^*_i(t)$: after this procedure we move from end to the start within one segment.
  `Rescaling' refers to renormalize adjoint solutions at the interface between segments: after this procedure we move to another time segment.
  }
  \label{f:subscripts explain}
\end{figure}

% Define notations on subscripts
We first define some notations, as shown in figure~\ref{f:subscripts explain}.
Let $T$ be the time length of the entire trajectory, and $K$ the total number of segments.
We denote the time span of the $i$-th segment  by $ [t_{i}, t_{i+1}] $, where $t_0=0, t_K = T$.
For quantities defined on a entire segment such as $\aW_i, \av^*_i$, $C_i$, $d_i$ and $a_i$,
their subscripts are the same as the segment they are defined on.
For quantities defined only at the interfaces between segments such as $Q_i, R_i$, $b_i$, $p_i$ and $\lambda_i$,
their subscripts are the same as the time point they are defined at.
Some of the notations are used immediately below, the others will be used in section~\ref{s:flowchart} and \ref{app:solve_NILSAS}.

% formula for W and v*
At time $t_i$, we perform QR factorization to $\aW_i(t) = [\aw_{i1}(t), \cdots \aw_{iM}(t)]$, 
which is a $m\times M$ matrix whose column vectors are homogeneous adjoints on segment $i$.
We use the $Q$-matrix, the matrix with orthonormal columns, as the terminal condition for $\aW_{i-1}$ on segment $i-1$.
More specifically, 
\begin{equation}\label{e:rescale_W}
  \aW_i(t_{i}) = Q_i R_i, \quad 
  \textnormal{and} \quad
  \aW_{i-1}(t_{i}) = Q_i \;.
\end{equation}

% rescale av*
At time $t_i$, we also rescale $\av^*_i(t)$, which is the particular inhomogeneous adjoint solution on segment $i$.
To do this, we subtract from $\av^*_i$ its orthogonal projection onto homogeneous adjoint solutions.
More specifically, 
\begin{equation} \label{e:rescale_v*}
  p_i := \av^{*}_i(t_{i}) - Q_i b_i \,,
  \textnormal{ where } 
  b_i = Q_i ^T \av^{*}_i(t_{i}) \,,
  \textnormal{ and } 
  \av^*_{i-1}(t_{i})=p_i \,.
\end{equation}
This rescaling maintains the continuity of the affine space $\av_i^{*}+span\{\aw_{ij}\}_{j=1}^M$ across different segments.

% give the formula for \av
The continuity of affine space allows us to impose continuity condition for $\av_i$, which is the adjoint shadowing direction on segment $i$.
On each segment,  $\av_i = \av^*_i + \aW_i a_i$ for some $a_i\in\R^M$.
The continuity condition can now be expressed via a relation between $a_i$ and $a_{i-1}$:
\begin{equation}
  \av^*_i(t_i) + \aW_i(t_i) a_i = \av^*_{i-1}(t_i) + \aW_{i-1}(t_i) a_{i-1} \;.
\end{equation}
Apply equation~\eqref{e:rescale_W} and~\eqref{e:rescale_v*}, cancel $\av^*_i(t_i)$ on each side, we get:
\begin{equation} 
  Q_i R_i a_i = -Q_i b_i + Q_i a_{i-1}
\end{equation}
Since $Q_i$ has orthonormal columns, $Q_i^TQ_i=I\in \R^{M\times M}$. 
Multiplying $Q_i^T$ to the left of both sides, we have the continuity condition for $\av$:
\begin{equation}
  a_{i-1} = R_i a_i + b_i \;. 
\end{equation}

\section{The NILSAS algorithm}

\subsection{Procedure list of the algorithm}\label{s:flowchart}

% what solver we need
Now we give a procedure list of the NILSAS algorithm.
To start with, we need to have an inhomogeneous adjoint solver and a homogeneous adjoint solver, both can take arbitrary terminal conditions.
The inhomogeneous adjoint equation we solve in NILSAS has right-hand-side $-J_u$, which is the same as many existing adjoint solvers.
Hence for inhomogeneous adjoint solvers in NILSAS, we only need to change existing solvers to be able to take arbitrary terminal conditions.
For homogeneous adjoint solvers, we only need to further change the right-hand-side of existing solvers to zero.

% other data
We provide the following data to NILSAS: 
1) the number of homogeneous adjoint solutions, $M\ge\mus+1$, where $\mus$ is the number of unstable CLVs; 
2) the total number of segments, $K$; 
3) for convenience, we assume that the length of all time segments are the same, denoted by $\Delta T$.
The  total time length is determined by $T =K \Delta T$.
Moreover, in the procedure list below, inner products are written in matrix notations, and by default vectors are in column forms.

\begin{enumerate}

  \item Integrate the primal system for sufficiently long time before $t=0$ so that $u(t=0)$ is on the attractor.

  \item Compute the trajectory $u(t), t\in [0,T]$, by integrating the primal system.

  \item Generate terminal conditions for $\aW_{i}$ and $\av^*_i$ on the last segment $i=K-1$:
  \begin{enumerate}
    \item Randomly generate a $m\times M$ full rank matrix, $Q'$.
      Perform QR factorization: $Q_K R_K = Q'$.
    \item Set $p_{K} = 0$.
  \end{enumerate}

  \item Compute $\aW_i$ and $v^*_i$ on all segments.
  For $i=K-1$ to $i=0$ do:
  \begin{enumerate}
    \item 
      To get $\aW_i(t)$, whose columns are homogeneous adjoint solutions on segment $i$, solve:
      \begin{equation}
        \dd {\aW_i}t + f_u ^T \aW_{i} = 0, \quad \aW_{i}(t_{i+1}) = Q_{i+1} \,.
      \end{equation}
      To get $\av^*_i(t)$, solve the inhomogeneous adjoint equation:
      \begin{equation}
        \dd {\av^*_i}t + f_u ^T \av^*_i = -J_u \,,\quad \av^*_i (t_{i+1}) = p_{i+1} \,.
      \end{equation}

    \item Compute the following integrations.
      \begin{equation} \label{e:covariant matrix} \begin{split}
        &C_i = \int_{t_{i}}^{t_{i+1}} \aW_i^T \aW_i dt \,, \quad
        d^{wv^*}_i = \int_{t_{i}}^{t_{i+1}} \aW_i^T \av^* dt \,, \\
        &d^{wf}_i = \int_{t_{i}}^{t_{i+1}} \aW_i^T f dt \,, \quad
        d^{v^*f}_i = \int_{t_{i}}^{t_{i+1}} \av^{*T} f dt \,, \\
        &d^{wf_s}_i = \int_{t_{i}}^{t_{i+1}} \aW_i^T f_s dt \,, \quad
        d^{v^*f_s}_i = \int_{t_{i}}^{t_{i+1}} \av^{*T} f_s dt \,, \\
        &d^{J_s}_i = \int_{t_{i}}^{t_{i+1}} J_s dt \,,
      \end{split} \end{equation}
      where $d^{wv^*}_i$, $d^{wf}_i$, $d^{wf_s}_i\in \R^M$; $d^{v^*f}_i$, $d^{v^*f_s}_i$, $d^{J_s}_i\in\R$;
      $C_i\in \R^{M\times M}$ is the covariant matrix.

    \item Orthonormalize homogeneous adjoint solutions via QR factorization:
      \begin{equation}\label{e:Ri}
        Q_i R_i = \aW_i(t_{i})
      \end{equation}

    \item Rescale the inhomogeneous adjoint solution using $Q_i$: 
      \begin{equation}\label{e:bi}
        p_i = \av^*_i(t_{i}) - Q_i b_i \,,\quad
        \textnormal{where } b_i = Q_i^T \av^{*}_i(t_{i}) \,.
      \end{equation}
  \end{enumerate}

\item Compute the adjoint shadowing direction $\{\av_i\}_{i=0}^{K-1}$.
  \begin{enumerate}
    \item Solve the NILSAS problem on multiple segments:
      \begin{equation} \label{e:NILSAS_on_multiple_segments}	\begin{split}
        &\min_{a_0,\cdots,a_{K-1} \in \R^M} \sum_{i=0}^{K-1} \frac 12 (a_i)^T C_i a_i +  (d^{wv^*}_i)^T a_i, \quad  \mbox{s.t.}\\
        &\;\;a)\;\; a_{i-1} = R_i a_i + b_i \,, \quad  i=1,\cdots,K-1 \;,\\
        &\;\;b)\;\; \sum_{i=0}^{K-1} (d^{wf}_i)^T a_i + \sum_{i=0}^{K-1} d^{v^*f}_i = 0 \;.
      \end{split}\end{equation}
      This is a least squares problem in $\{a_i\}_{i=0}^{K-1} \subset \R^M$.
      In \ref{app:solve_NILSAS} we suggest one way to solve this problem.

    \item On each time segment $i$, $\av_i $ is given by
      \begin{equation} \label{e:y_i}
        \av_i(t) = \av^*_i(t) + \aW_i(t) a_i .
      \end{equation}
  \end{enumerate}

  \item Compute the derivative by:
    \begin{equation} \label{e:dJds_multiple_segments}
      \frac{d\avg{J}}{ds} \approx
      \frac 1T \sum_{i=0}^{K-1} 
      \int_{t_i}^{t_{i+1}} \left( \av_i^T f_s  +J_s \right) dt
      = \frac 1T \sum_{i=0}^{K-1} \left( d^{v^*f_s}_i + a_i^T d^{wf_s}_i + d^{J_s}_i \right)
    \end{equation}
\end{enumerate}

\subsection{Remarks about NILSAS}\label{s:remarks on NILSAS}

\subsubsection{Miscellaneous}\label{s:remarks misc}

% remark: what to not store
We remark that if we are not interested in obtaining $\av(t)$ for all $t$, 
there is no need to store adjoint solutions $\aW$ and $\av^*$ in computers, which is typically lots of data.
To compute sensitivity, we only need to store $d^{wv^*}_i$, $d^{wf}_i$, $d^{wf_s}_i$,
$d^{v^*f}_i$, $d^{v^*f_s}_i$, $d^{J_s}_i$, $C_i$ given in equation~\eqref{e:covariant matrix},
$R_i$ given in equation~\eqref{e:Ri}, and $b_i$ given in equation~\eqref{e:bi}.

% remark: taking snapshots
Moreover, when computing quantities in equation~\eqref{e:covariant matrix}, 
we can estimate the integration using particular values of the integrands evaluated at several snapshots,
to further reduce the storage management cost.
For example, similar to \cite{Blonigan_2017_adjoint_NILSS}, we can estimate $C_i$ by the terminal value of $W_i$, which is $Q_i$,
and thus $C_i=I\in\R^{M\times M}$.
We suggest further research be done to determine which estimation is best practice.

% remark: cost of NILSAS
NILSAS has the benefit of typical adjoint algorithms,
that is, for a new parameter $s$, $\av$ does not change, 
so we only need to give new $f_s$, $J_s$, and recompute equation~\eqref{e:covariant matrix} and \eqref{e:dJds_multiple_segments}.
Hence the extra cost for a new parameter is only performing an $L^2$ inner product, which is negligible in comparison with the total cost of the algorithm.
More specifically, assume there are $n$ parameters $s=[s_1,\cdots,s_n]$, we can define 
\begin{equation}\begin{split}
  \dd {\avg{J}} s &= [\pp {\avg{J}} {s_1}, \cdots, \pp {\avg{J}} {s_n} ] \in R^{1\times n} \,;\\
  J_s &= [\pp J {s_1}, \cdots, \pp J {s_n}]  \in R^{1\times n}  \,;\\
  f_s &= [\pp f {s_1}, \cdots, \pp f {s_n}]  \in R^{m\times n}  \,.\\
\end{split}\end{equation}
With these definitions, the NILSAS algorithm, in particular equation~\eqref{e:covariant matrix} and \eqref{e:dJds_multiple_segments},
extend to several parameters with almost no extra cost.
An extreme example is where $f_s$ is unknown a priori and we can now use $\av$ to design an optimal control, $f_s$.

  % non-uniform system
  The assumptions in theorem~\ref{thm:flow} are made for theoretically proving the unique existence of adjoint shadowing directions and convergence of NILSAS.
  In practice, it is possible that NILSS/NILSAS are still valid on a chaotic system which fails these assumptions.
  For example, the 3D cylinder flow we investigate later in this paper has at least two neutral CLVs, 
  corresponding to translations in time and in the span-wise directions, due to the periodic boundary condition.
  In fact, 
  in \cite{Ni_CLV_cylinder} we also showed that the smallest angle between tangent CLVs depends on meshes and may fall below a threshold value:
  this further violates our assumptions.
  However, we did found the trend that angles between tangent CLVs gets larger when their indices are further apart: 
  this property is related to hyperbolicity, but has not been well investigated yet.
  As we shall see, both NILSS and NILSAS compute correct sensitivities on this 3D flow.
  The generality of shadowing methods is as suggested by the chaotic hypothesis \cite{gallavotti_chaotic_hypothesis_1995,Gallavotti:2008},
  that is, theoretical tools may still valid for non-uniform hyperbolic chaotic systems, 
  even though those tools can only be rigorously proved with a stricter assumption.
  We do not expect NILSS and NILSAS be valid for all chaotic systems;
  however, they are valid somewhere beyond our current assumptions.
  %For example, we suspect that the trend of angles between CLVs we found in \cite{Ni_CLV_cylinder} is enough for NILSS/NILSAS to work.
  We call for more research to identify the limit of shadowing methods, especially in real-life problems.

% remark about diffeomorphism
In \ref{s:discrete systems}, we discuss in detail NILSAS for discrete systems, more specifically, hyperbolic diffeomorphisms.
Adjoint shadowing directions for diffeomorphisms were also defined in \cite{Ni_adjoint_shadowing}.
Because of the absence of the neutral subspace, the NILSAS algorithm for hyperbolic diffeomorphisms is easier than flows.
To obtain NILSAS for diffeomorphisms, we no longer compute $d^{wf}_i$, $d^{v^*f}_i$, 
and no longer impose the second constraint in the NILSAS problem.
Of course, we should change integrations to summations, and adjoint equations to their discrete counterparts.

\subsubsection{Number of homogeneous adjoint solutions}\label{s:M}

Since the number of homogeneous adjoint solutions should be strictly larger than then number of unstable adjoint CLVs, 
which equals the number of unstable tangent CLVs, 
we first discuss the number of unstable CLVs, about which there are two questions: 
(1)  whether the absolute number can be large; (2) whether the number is significantly lower than the dimension of the system.
We are interested in (1) because we want to estimate the cost of NILSAS.
We are interested in (2) because we want to determine whether computational efficiency can benefit from the non-intrusive formulation, which restricts minimization to unstable subspaces.
Roughly, the efficiency improvement due to the non-intrusive formulation is proportional to the ratio of the system dimension to unstable subspace dimension.

First, the absolute number of unstable CLVs can be large and the cost of NILSAS increases.
We think maybe this is the price to pay for chaos, 
that is, for more chaotic systems, numerical methods should be more expensive, not only for NILSS/NILSAS, 
but also for other common methods such as computing long-time averages, which should take longer time to converge for more chaotic systems.
Second, in a recent paper \citep{Ni_CLV_cylinder}, based on observations on flow past a 3D cylinder, we conjectured that for open flows, CLVs active in the freestream or less turbulent regions are stable.
At least for these open flows, where there are large areas of freestream, a large fraction of CLVs should be stable.
For these cases, we can benefit from the non-intrusive formulation by restricting minimization to unstable subspaces.
On the other hand, there are cases where significant part of all CLVs are unstable, such as Hamiltonian systems.
For such cases the non-intrusive formulation does not help reduce the computational cost.
The other possible case in fluid mechanics might be wall bounded turbulence,
where for higher Reynolds number there are more unstable CLVs, but also more CLVs since finer meshes should be used, 
and we do not know yet if, in the limit, a significant fraction of CLVs can be unstable.

We provide some examples in computational fluid mechanics on the number of unstable CLVs and the dimension of the system.
For a 2D incompressible channel flow over a backward facing step at Reynolds number $Re=2.5\times 10^4$, 
there are $13$ unstable CLVs in a system of dimension $4\times 10^4$ \cite{Ni_NILSS_JCP}.
For a 2D NACA 0012 airfoil at Mach number $Ma=0.2$, angle of attack $20\deg$, $Re=2400$,
there are less than 5 unstable CLVs for different implementations with system dimension
ranging from $7\times 10^3$ to $8\times 10^5$ \cite{Fernandez_LE_discretization}.
For a 3D turbulent channel flow with $Ma=0.3$ and $Re_\tau = 180$ on a domain of size $4\pi\times 2\times 2\pi$, 
there are about $1.5\times 10^3$ unstable CLVs out of a system of dimension $2.2\times 10^6$ \cite{Blonigan_channelLE}.
For a 3D weakly turbulent flow over a cylinder at $Re=5.2\times 10^2$, 
there are $20$ unstable out of $1.9\times 10^6$ \cite{Ni_CLV_cylinder}.
To conclude, we believe that although the cost of NILSS and NILSAS can be high, 
for many cases, the non-intrusive formulation is beneficial for computational efficiency.

In our procedure list we listed $M$ in the setting of NILSAS and required it strictly larger than $m_{us}$, the number of unstable CLVs.
How should we know $m_{us}$ before running NILSAS? And what if we chose initial $M$ smaller than required?
First, the number of unstable modes is roughly positively related to how chaotic the flow is.
This is not a rigorous criterion but readers can look at some test cases to have a rough sense.
But there is not any precise method that allows us to know the exact number at the first glance.
Second, even if we started with an insufficient $M$, we can add adjoint solutions inductively in NILSAS,
rather than recomputing everything all over again.
More specifically, in the NILSAS problem in equation~\eqref{e:NILSAS_on_multiple_segments} 
and the sensitivity formula in equation~\eqref{e:dJds_multiple_segments},
say we want to add $k$ more adjoint solutions,
then coefficients arrays $d^{wv^*}_i$, $d^{wf}_i$, $d^{wf_s}_i$ and $b_i$,
should be augmented by $k$ more entries, 
while the old coefficient arrays are not changed inside the new arrays;
similarly, the coefficient matrices $C_i$, $R_i$ should be augmented by $k$ rows and $k$ columns.

The headache of choosing an initial $M$ is further relieved by the fact that adjoint solutions can be more efficiently computed in batches.
Within each segment, we can accelerate NILSAS by taking advantage of the fact that all adjoint solutions, 
both homogeneous and inhomogeneous, use the same Jacobian $f_u$.
If the numerical integration is vectorized, 
we can integrate all adjoint solutions simultaneously without repeatedly loading $f_u$ into the computer CPU, 
which is the most time-consuming procedure in the numerical integration.
At each time step, instead of several matrix-vector products, we can perform one matrix-matrix products,
where the second matrix is composed of several adjoint solutions;
then we add the right-hand-side to the inhomogeneous adjoint solution.
For example, for a 4th order IEDG solver, the marginal cost for one more adjoint solution can be only, say 0.037, 
of the first adjoint solution.
In this scenario we should start NILSAS and then add adjoint solutions by batches on the order of $1/0.037\approx 27$ adjoint solutions per batch.
This should be further faster than adding adjoint solutions one by one.
\footnote{The ideas of taking advantage of vectorized integration 
  and the estimation on IEDG were both given during private discussion by Pablo Fernandez, 
  who co-authored with us on finite difference NILSS (FD-NILSS), see arXiv:1711.06633.}

\subsubsection{Other settings of NILSAS}

It is required by the algorithm that we run primal system long enough before the main part of NILSAS, so that our initial condition is on the attractor.
In general, we can not know very well what is `long enough' before we do any computations, 
and this run-up time is determined empirically as the time when the flow field starts to repeat itself.
In a typical scenario, we would run a primal simulation before taking interest in any sensitivities.
When running that primal simulation, there is the same question of when we reach the stage that enough long-time behavior has been captured:
typically this is indicated by that several objective functions began to oscillate around some averaged values.

We should also determine the time length $T$ on which we run NILSAS.
In practice $T$ is determined empirically as the time when the sensitivity computed by NILSAS converges to within the uncertainty bound we desire.
However, there is one caveat that the adjoint solutions are computed backwards in time.
Now if we find $T$ insufficient, we can not add time after $T$ without recomputing all adjoint solutions, since integration adjoint solutions forward in time is very unstable.
Rather, we should add time before our current trajectory.
In practice, we should run our primal simulation till enough long-time behavior has been captured, then start computing adjoint solutions from the end of that primal trajectory.
We have found that typically NILSAS requires a shorter trajectory to compute sensitivity than that required to reflect average behavior.

Then we discuss the choice of segment length $\Delta T$.
Similar to NILSS, $\Delta T$ is determined by that within one segment, the leading adjoint CLV does not dominate the $M$-th adjoint CLV.
This is because otherwise we would have covariant matrices, $\{C_i\}_{i=0}^{K-1}$, with small condition number, 
which would lead to eventually the poor condition of the NILSAS problem in equation~\eqref{e:NILSAS_on_multiple_segments}.
We recommend $\Delta T (\lambda_1-\lambda_M)$ to be $O(1)$, in which case within one segment,
the leading CLV would grow to be about $e^1=2.7$ time larger than the $M$-th CLV.

A related question is that if the leading LE is large and we select a small $\Delta T$, will the cost of frequent rescaling offset the cost reduction due to non-intrusive formulation?
First, as we discussed in section~\ref{s:M}, there are a lot of fluid systems whose CLVs are mostly stable, in which case the non-intrusive formulation is beneficial.
Second, our understanding is that the numerical methods should have smaller time steps for more chaotic systems, to capture accurate motions on all scales.
Hence one segment, although is shorter in physical time, may still contain many small time steps.
As a result, the rescaling may not be more frequent for more chaotic systems. 
Again, we call for more research on numerical schemes and LE spectrum, 
especially for systems other than fluid or extremely chaotic systems.

\subsubsection{Comparison with other shadowing algorithms}

% compare with LSS
There are currently several variants of NILSS \cite{Ni_NILSS_JCP},
such as the Finite-Difference NILSS (FD-NILSS) \cite{Ni_fdNILSS} and discrete adjoint NILSS \cite{Blonigan_2017_adjoint_NILSS}.
NILSAS, as well as these NILSS variants, bears part of the merit of `non-intrusive' formulation, that is, in comparison to LSS,
the minimization problems in these algorithms are constrained to the unstable subspaces.
Hence, for many real-life problems, where the unstable subspaces have significantly lower dimension than the dynamical systems,
these algorithms should be significantly faster than LSS.

% compare with other non-intrusive algorithms
We compare NILSAS with variants of NILSS in table~\ref{t:compare}.
In particular, we want to compare in more detail the two adjoint algorithms: discrete adjoint NILSS versus NILSAS.
%Adjoint discrete NILSS requires both tangent and adjoint solvers, and requires programming a new non-zero right-hand-side for adjoint equations,
%whereas NILSAS requires only adjoint solvers, with either has zero right-hand-side, 
%or the conventional right-hand-side as implemented in most existing adjoint solvers: hence NILSAS should be easier to implement.
NILSAS should be easier to implement than the discrete adjoint NILSS since it does not require tangent solvers, 
and requires less modification to existing adjoint solvers. 
Furthermore, unlike discrete adjoint NILSS, 
NILSAS does not explicitly depend on the fact that inner-products between adjoint and tangent homogeneous solutions are constants:
since this property holds true only for analytic solutions but is typically false for numerical solutions,
we think NILSAS should be more robust to implementations of tangent and adjoint solvers, and should typically have better convergence.
We suggest more numerical comparison be done to compare the two methods.

\begin{table}[ht] \centering
  \makebox[\linewidth]{
  \begin{threeparttable} 
  \begin{tabular}{l | c c c c} 
    \hline
                                                              & NILSS & FD-NILSS  & D.A. NILSS  & NILSAS \\
    \hline
    1. needs prm solvers                                      & yes   & yes       & yes         & yes\\
    2. needs tan solvers                                      & yes   & no        & yes         & no \\
    3. needs adj solvers                                      & no    & no        & yes         & yes\\
    \makecell[cl]{4. cost increases with \\parameter numbers} & yes   & yes       & no          & no\\
    \makecell[cl]{5. cost increases with \\objective numbers} & no    & yes\tnote{$\dagger$}       & yes         & yes\\
    \makecell[cl]{6. cost for 1 parameter \\and 1 objective} & 
          \makecell{ 1 prm\\+ 1 ihm tan\\+ (M-1) hm tan}&(M+1) prm 
          &\makecell{ 1 prm\\+1 ihm adj \\+ (M-1) hm tan}&\makecell{ 1 prm\\+1 ihm adj \\+ M hm adj}\\
    \hline
  \end{tabular}
  \begin{tablenotes}
  \item[$\dagger$] 
    The cost of FD-NILSS may not increase with objective numbers if $J_u$ and $J_s$ can be provided without finite difference procedure, 
    which may not be true if working with only typical primal solvers.
  \end{tablenotes}
  \end{threeparttable}}
  \caption{Comparison of NILSAS with NILSS, Finite Difference NILSS (FD-NILSS) and discrete adjoint (D.A) NILSS.
  Here `prm', `tan', `adj', `ihm' and `hm' are short for primal, tangent, adjoint, inhomogeneous and homogeneous, respectively.
M is a number strictly larger than the number of unstable CLVs.}
  \label{t:compare}
\end{table}

\section{Applications}

\subsection{Application on Lorenz 63 system}

% introduces the system
In this subsection we apply NILSAS to the Lorenz 63 system as an illustration.
\footnote{The python code used for this section is at: https://github.com/niangxiu/nilsas.}
Lorenz 63 is an ordinary differential equations system with three states $u = [x, y, z]$:
\begin{equation}
  \dd{x}{t} = \sigma(y-x),
  \quad \dd{y}{t}=x(\rho-z)-y,
  \quad \dd{z}{t} = xy- \beta z.
\end{equation}
We fix $\beta=8/3$.
This system models the heat transfer in a fluid layer heated from below and cooled from above.
In particular, $x$ is the convection rate, $y$ the horizontal temperature variation, and $z$ the vertical temperature variation.
The parameters $\sigma$ and  $\rho$ are proportional to the Prandtl number and Rayleigh number.
We select the instantaneous objective function as $J(u) = z$, and hence our objective $\avg{J}$ is the averaged vertical temperature variation.

% settings for numerical integration and NILSAS
The primal system and adjoint equations are integrated via the explicit time-stepping scheme:
\begin{equation} \begin{split}
  u_{k+1} &= u_k + f(u_k) \Delta t \\
  \aw_{k} &= \aw_{k+1}  + f_u(u_k)^T \aw_{k+1} \Delta t\\
  \av^*_{k} &= \av^*_{k+1}  + f_u(u_k)^T \av^*_{k+1} \Delta t + J_u(u_k) \Delta t
\end{split} \end{equation}
where the subscript $k$ denotes the time step number in numerical integration.
The time step size is $\Delta t = 0.001$.
For NILSAS, time segment length is $\Delta T=0.2$, thus there are 200 time steps per segment.

% find M
We want to determine the number of unstable CLVs for the Lorenz system.
The Lyapunov exponents, $\lambda_1, \lambda_2, \lambda_3$, satisfy the following constraints \cite{bovy2004lyapunov}:
\begin{equation}
  \lambda_1 + \lambda_2 + \lambda_3 = -(1+\sigma+\beta) < 0 \;.
\end{equation}
Moreover, one of these exponents corresponds to the neutral CLV so it is zero.
Hence there is at most one positive exponent, so in NILSAS we set the number of homogeneous solutions $M=2$.

\begin{figure}[ht]
\centering
  \begin{subfigure}{0.47\textwidth}
    \includegraphics[width=\textwidth] {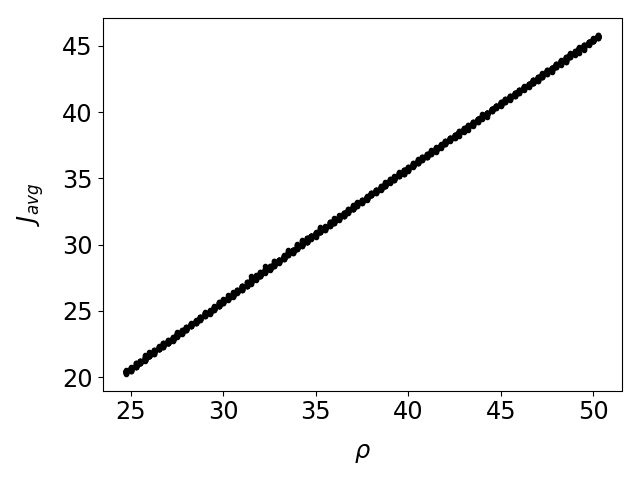}
    \caption{For each value of $\rho$, $\avg{J}$ is computed 20 times on randomly initialized trajectories of length $100$.}
  \end{subfigure}
  \hfill
  \begin{subfigure}{0.47\textwidth}
    \includegraphics[width=\textwidth] {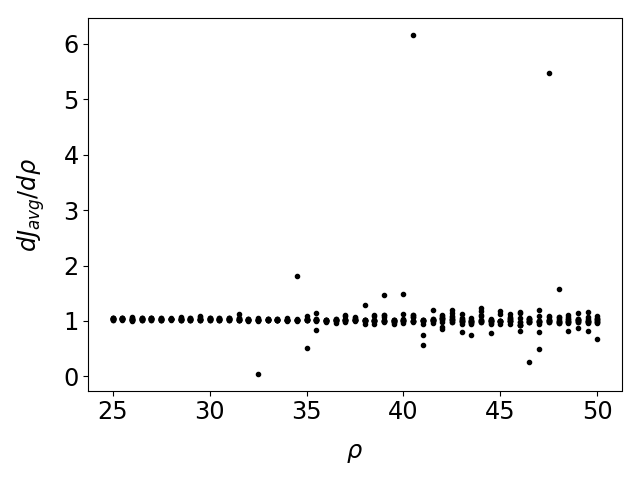}
    \caption{For each $\rho$, $\partial\avg{J}/\partial\rho$ is computed 
      10 times by NILSAS on randomly initialized trajectories of length $40$.}
  \end{subfigure}
  \caption{$\avg{J}$ and $\partial\avg{J}/\partial\rho$ versus $\rho$ for the Lorenz 63 system. Here $\sigma = 10$ is fixed.}
  \label{f:rho J}
\end{figure}

% show results: change rho
We verify that NILSAS gives correct sensitivities by computing $\avg{J}$ and $\partial \avg{J}/ \partial \rho$ for different $\rho$, while fixing $\sigma=10$.
The Lorenz system has one quasi-hyperbolic strange attractor when $ 25\le\rho <31 $, and one non-hyperbolic attractor when $31 \le \rho \le 50$:
none of these cases strictly satisfies our uniform hyperbolic assumption.
As shown in figure~\ref{f:rho J}, as $\rho$ becomes larger, 
the system becomes non-hyperbolic, and the sensitivity results given by NILSAS begin to oscillate.
Nevertheless, NILSAS gives that $\partial\avg{J}/\partial{\rho}$ is approximately 1 for all $\rho$, 
which matches the trend between $\avg{J}$ and $\rho$:
this again shows that NILSAS can be effective for systems not satisfying assumptions of theorem~\ref{thm:flow}, 
as we discussed in section~\ref{s:remarks misc}.

\begin{figure}[ht]
\centering
  \begin{subfigure}{0.47\textwidth}
    \includegraphics[width=\textwidth] {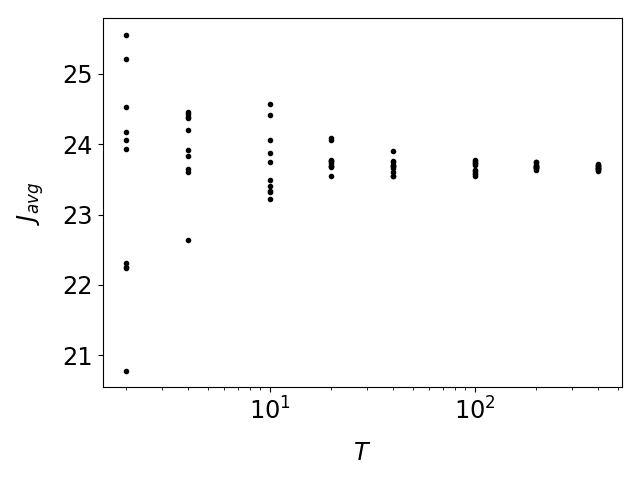}
    \caption{For each value of trajectory length $T$, $\avg{J}$ is computed 10 times on randomly initialized trajectories.}
  \end{subfigure}
  \hfill
  \begin{subfigure}{0.47\textwidth}
    \includegraphics[width=\textwidth] {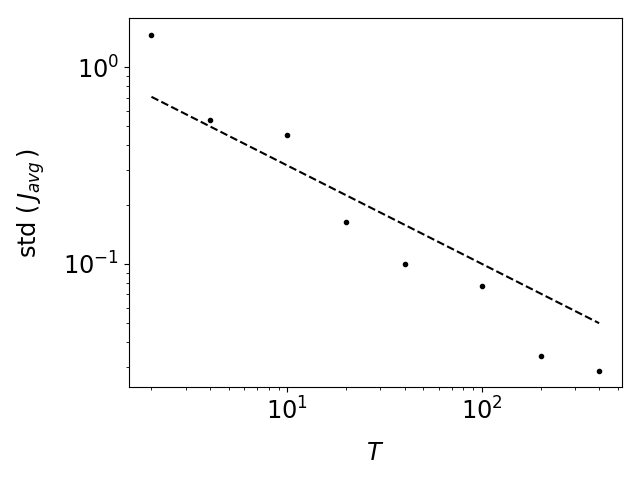}
    \caption{The sample standard deviation of the 10 $\avg{J}$'s computed at each $T$. 
      The dashed line is $T^{-0.5}$.}
  \end{subfigure}
  \caption{Convergence of the averaged objective $\avg{J}$ with respect to the trajectory length $T$.
    Here $\rho=28$ and $\sigma = 10$ are fixed.}
  \label{f:T J}
\end{figure}

\begin{figure}[ht]
\centering
  \begin{subfigure}{0.47\textwidth}
    \includegraphics[width=\textwidth] {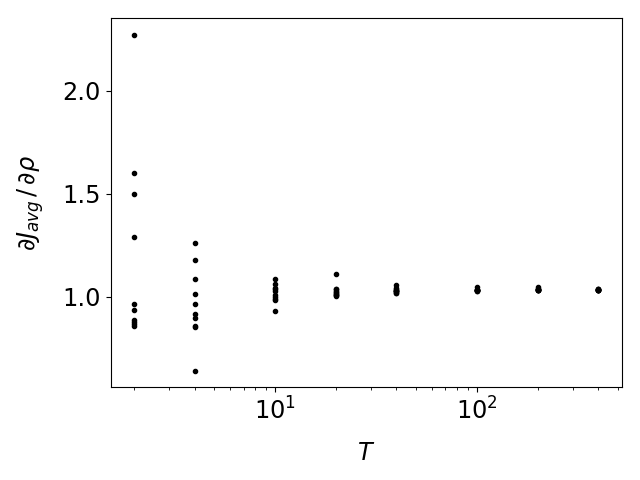}
    \caption{For each value of $T$, $\partial\avg{J}/\partial\rho$ is computed by NILSAS 10 times on randomly initialized trajectories.}
  \end{subfigure}
  \hfill
  \begin{subfigure}{0.47\textwidth}
    \includegraphics[width=\textwidth] {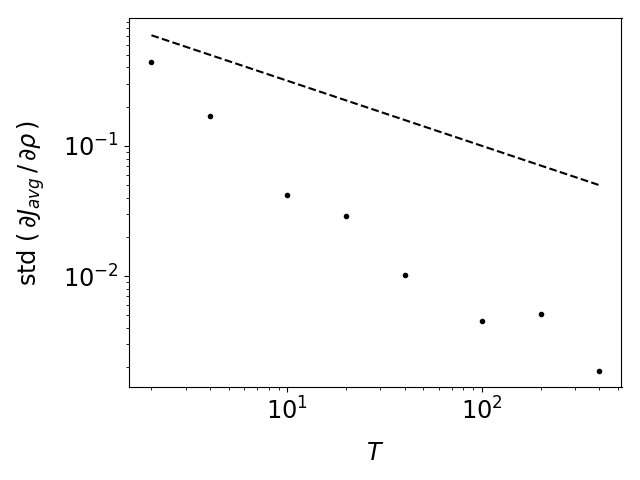}
    \caption{The sample standard deviation of the 10 $\partial\avg{J}/\partial\rho$'s computed at each $T$.
      The dashed line is $T^{-0.5}$.}
  \end{subfigure}\\
  \vspace*{6mm}
  \begin{subfigure}{0.47\textwidth}
    \includegraphics[width=\textwidth] {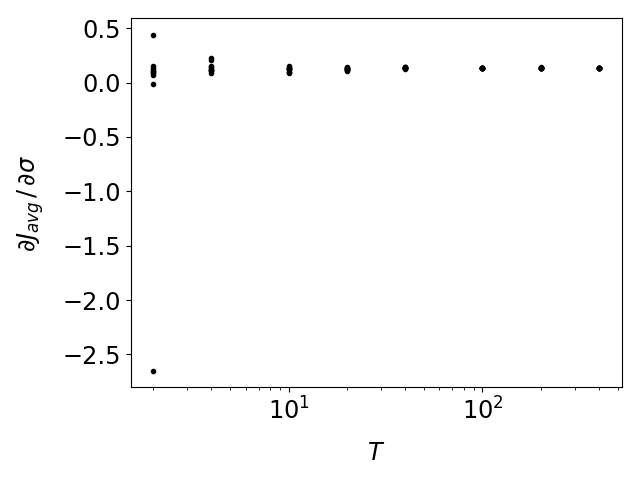}
    \caption{For each value of $T$, $\partial\avg{J}/\partial\sigma$ is computed by NILSAS 10 times on randomly initialized trajectories.}
  \end{subfigure}
  \hfill
  \begin{subfigure}{0.47\textwidth}
    \includegraphics[width=\textwidth] {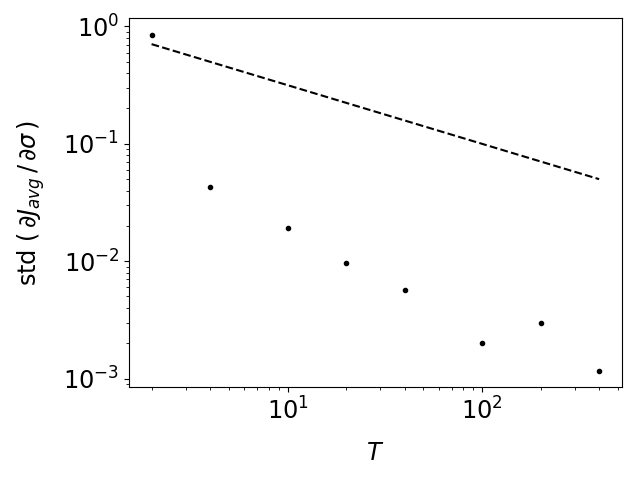}
    \caption{The sample standard deviation of the 10 $\partial\avg{J}/\partial\sigma$'s computed at each $T$.
      The dashed line is $T^{-0.5}$.}
  \end{subfigure}
  \caption{Convergence of sensitivities computed by NILSAS with respect to the trajectory length $T$.
    Here $\rho=28$ and $\sigma = 10$ are fixed.}
  \label{f:T dJdrho dJdsigma}
\end{figure}

% show results: convergence of J ~ T
Then we show that both $\avg{J}$ and the sensitivities computed by NILSAS converge as the trajectory length $T$ gets larger,
while fixing $\sigma=10$ and $\rho=28$.
Figure~\ref{f:T J} shows that the standard deviation of $\avg{J}$ reduces at the rate of $T^{-0.5}$.
Figure~\ref{f:T dJdrho dJdsigma} shows that the sensitivities computed by NILSAS, 
with respect to both $\rho$ and $\sigma$, converge faster than the rate of $T^{-0.5}$.

% show results: convergence of gradient ~ T

\begin{figure}[ht]
  \centering
  \includegraphics[width=0.75\textwidth] {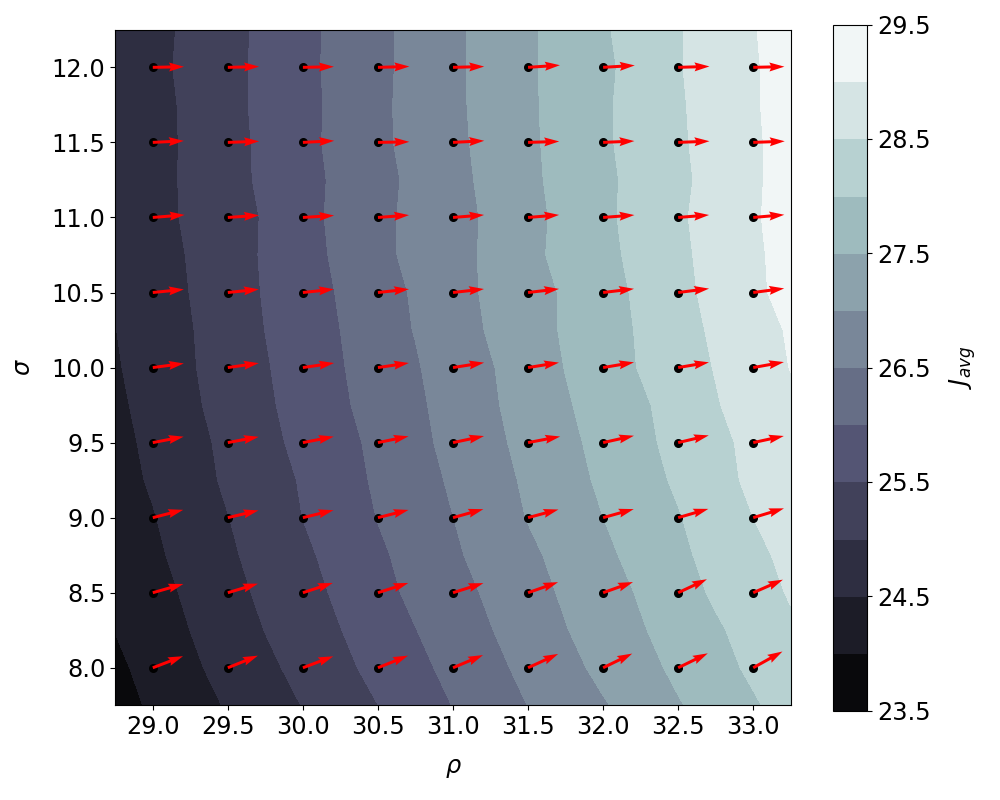}
  \caption{Gradients computed by NILSAS. 
    The contour is of $\avg{J}$ with respect to $\rho$ and $\sigma$, 
    and red vectors are gradients.
    Here $\avg{J}$'s are averaged over 20 randomly initialized trajectories of length $100$,
    while gradients computed by NILSAS are averaged over 10 randomly initialized trajectories of length $40$.
    The vector length is $0.2$ times the gradient norm.
    NILSAS computes one gradient, composed of two sensitivities to two parameters, in one run.}
  \label{f:contour}
\end{figure}

% show results: gradient contour
NILSAS computes sensitivities with respect to multiple parameters with almost no additional cost, 
since the adjoint shadowing solution $\av$ does not depend on the choice of parameters.
Figure~\ref{f:contour} illustrates the contour of $\avg{J}$ with respect to $\rho$ and $\sigma$, 
and the gradient, $[\partial \avg{J} /\partial \rho, \partial \avg{J} /\partial \sigma]$, 
is computed by NILSAS.
Since we use the same length unit for both parameters, gradients should be perpendicular to the level sets of the objective:
this is indeed the case, and it shows NILSAS gives correct gradient information.

\begin{figure}[ht]
  \centering
  \includegraphics[width=0.95\textwidth] {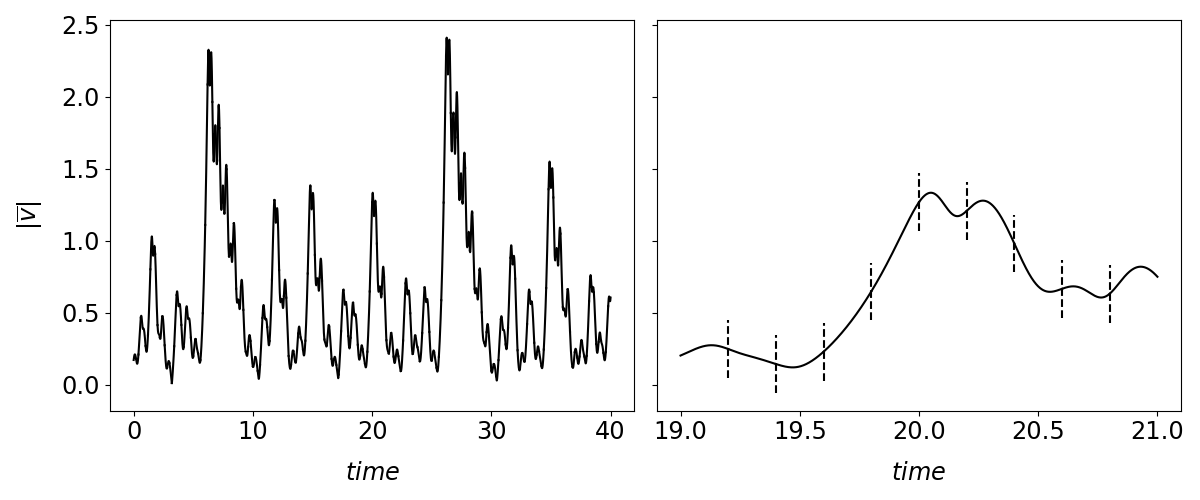}
  \caption{Norm of the adjoint shadowing direction computed by NILSAS for the Lorenz system, with $\rho=28$ and $\sigma = 10$. 
    Left: plot on the entire trajectory time span.
    Right: zoom onto time span from 19 to 21.
    The vertical dashed lines marks different time segments.
    }
  \label{f:av_norm}
\end{figure}

% show the adjoint shadowing direction
Finally, we draw the norm of an adjoint shadowing direction in figure~\ref{f:av_norm}.
As we can see from the left plot,
the norm of the adjoint shadowing direction does not grow exponentially,
satisfying the third property of adjoint shadowing directions.
Moreover, as shown in the right plot,
the adjoint shadowing direction computed by NILSAS is continuous.
This shows that our dividing trajectory technique indeed allows us to recover a continuous adjoint shadowing direction.

\subsection{Application on a weakly turbulent flow past a three-dimensional cylinder}

\begin{figure}[htb!]
    \centering
    \includegraphics[width=0.6\linewidth]{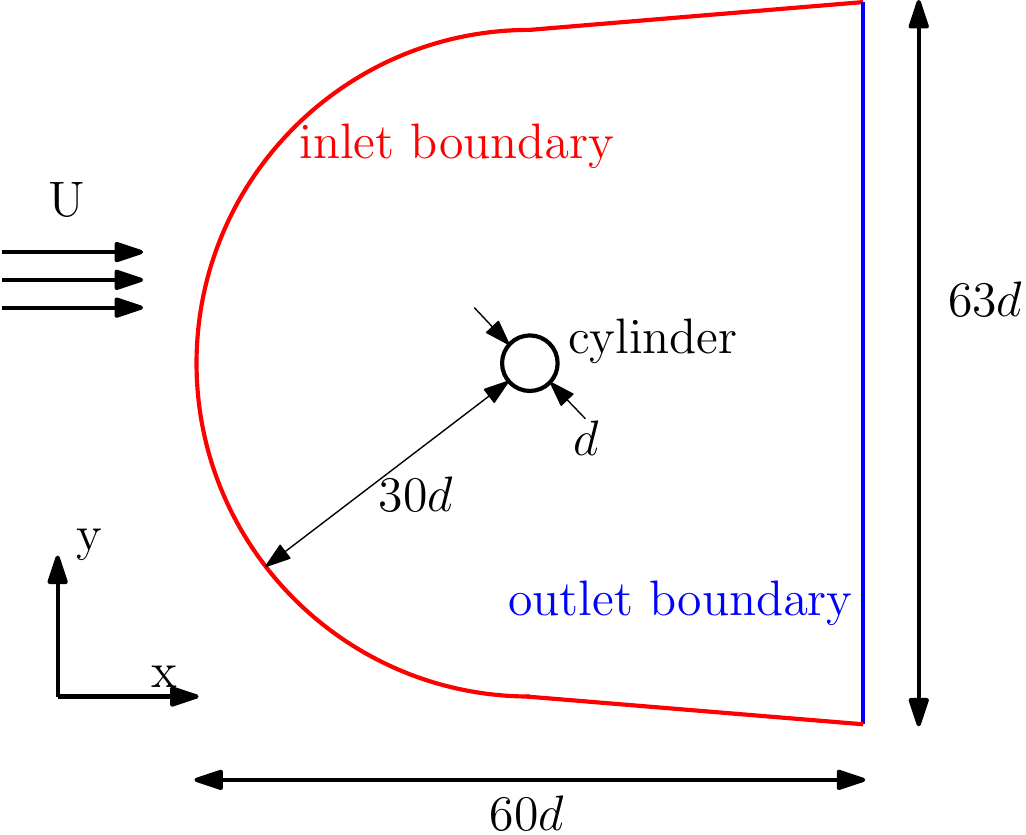}
    \caption{
     Geometry used in the simulation of a 3D flow past a cylinder. 
     The span-wise extent of the computational domain is 2d.}
    \label{f:cylinder_domain}
\end{figure}

% Description of subsonic flow over a 3D cylinder
In this subsection, we apply NILSAS to a 3D subsonic flow over a cylinder at Reynolds number $Re=1100$ and Mach number $Ma=0.093$.
\footnote{The flow solver, adFVM, used for this section is at: https://github.com/chaitan3/adFVM, the particular file that implements the NILSAS algorithm used in this case is apps/nilsas.py.}
The flow-wise length of the domain is $60d$, where $d=0.25 mm$  is the diameter of the cylinder. 
The Reynolds number is defined using the diameter of the cylinder and the density, velocity and viscosity of inflow. 
The span-wise extent, at $z = 2d$, is sufficient to capture most of the important flow features, like a turbulent wake and flow separation.
The front view of our fluid problem is shown in figure~\ref{f:cylinder_domain}.

% governing equations 
We use compressible Navier-Stokes equations with the ideal gas law approximating the thermodynamic state equation \cite{garnier2009large}.
The gas is assumed to be air.
More specifically, the governing equations are:
\begin{align}
\begin{split}
    &\frac{\partial \rho}{\partial t} + \nabla \cdot (\rho \mathbf{u}) = 0 \,, \\
    &\frac{\partial (\rho \mathbf{u})}{\partial t} + \nabla \cdot (\rho \mathbf{u}\mathbf{u}) + \nabla p = \nabla \cdot \sigma +  \,, \\
    &\frac{\partial (\rho E)}{\partial t} + \nabla \cdot (\rho E \mathbf{u} + p \mathbf{u}) = \nabla \cdot (\mathbf{u} \cdot \sigma + \alpha \rho \gamma \nabla e) \,,\\
    &\sigma = \mu (\nabla \mathbf{u} + \nabla \mathbf{u}^T) - \frac{2\mu}{3}(\nabla \cdot \mathbf{u})\mathbf{I} \,,
    \quad c = \sqrt{\frac{\gamma p}{\rho}} \,,\\
    &p = (\gamma - 1)\rho e \,,
    \quad e = E - \frac{\mathbf{u}\cdot \mathbf{u}}{2} \,.\\
\end{split}
\label{e:navier}
\end{align}
Here $\rho$ is the density, $\mathbf{u}$ is the velocity vector, $\rho E$ is the total energy,
$p$ is pressure, $e$ is internal energy of the fluid, $c$ is the speed of sound,
$\gamma = 1.4$ is the isentropic expansion factor and
$\mu$ is the viscosity field modeled using Sutherland's law for air 
\begin{equation}
    \mu = \frac{C_s T^{3/2}}{T + T_s}
\end{equation}
where $T_s = 110.4\,K$ and $C_s = 1.458 \times 10^{-6} {kg}/{m\,s\,\sqrt{K}}$.
$\alpha$ is the thermal diffusivity modeled using
\begin{equation}
    \alpha = \frac{\mu}{\rho Pr}
\end{equation}
where $Pr = 0.71$ is the Prandtl number.

\begin{figure}[htb!]
    \centering
    \includegraphics[width=0.8\linewidth]{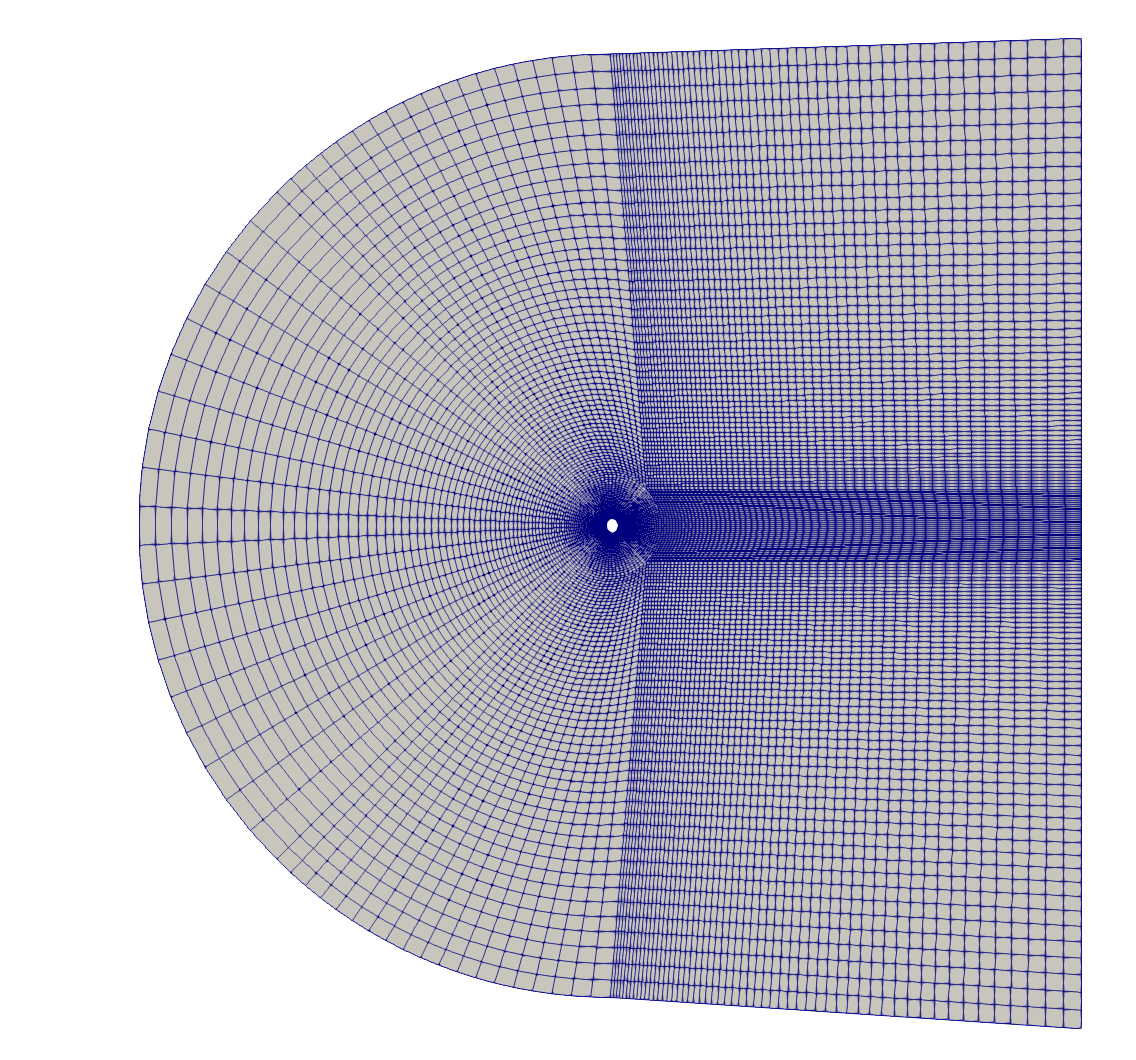}
    \caption{Front view of the mesh for the flow over cylinder problem.
    This is an unstructured hexahedral mesh with approximately $7\times 10^5$ cells, with $50$ cells in the span-wise direction.} 
    \label{f:cylinder_mesh}
\end{figure}

% Mesh
We use an unstructured hexahedral mesh with approximately $7\times 10^5$ cells, with $50$ cells in the span-wise direction.
The front view of our mesh is shown in figure~\ref{f:cylinder_mesh}.

% Numerical solver; adFVM
We use a second order finite volume method (FVM) \cite{leveque2002finite} for unstructured hexahedral meshes.
The central differencing scheme is used to interpolate cell averages of the flow solution onto faces of the mesh \cite{versteeg1995computational}. 
The numerical fluxes for the conservative flow variables are computed using the Roe approximate Riemann solver \cite{roe1981approximate}. 
An explicit time integration scheme, the strong stability preserving third order Runge-Kutta method \cite{macdonald2003constructing}, 
is used for time marching the numerical flow solution.
The size of the time step is determined using the acoustic Courant-Friedrichs-Lewy (CFL) condition \cite{courant1967partial},
and we choose our CFL number to be $1.2$.
The flow solver is implemented in Python using the adFVM \cite{talnikar2018} library,
which provides a high-level abstract application programming interface for writing efficient CFD applications.
The flow solver is parallelized using the Message Passing Interface (MPI) library.

% LES
We use implicit Large Eddy Simulations (LES) in our numerical simulation.
In an LES, the large scale eddies of the flow are resolved by the grid, 
while the contribution from the small scale eddies to the filtered Navier-Stokes equations 
are modeled using a sub-grid scale Reynolds stress model \cite{garnier2009large}.
In this paper, the numerical error of the discretization scheme serves as the LES model.
It has been shown that when using a relatively dissipative discretization method,
the numerical viscosity from the grid can be of the same order of magnitude as the sub-grid scale viscosity
\cite{moeng1989evaluation,fernandez2017subgrid},
and thus can be regarded as an implicit LES model.

% boundary condition
On the inlet boundary, we specify stagnation pressure and temperature, corresponding to a fixed Reynolds number $Re=1100$ 
and a Mach number which we choose to be the system parameter.
For the base case, we choose Mach number $Ma=0.093$.
Periodic boundary condition is used in the span-wise direction. 
The surface of the cylinder is maintained at a constant temperature of $300 K$.
Static pressure of 1 atmosphere unit is prescribed on the outlet boundary.

\begin{figure}[htb!]
    \centering
    \includegraphics[trim=0cm 10cm 0cm 10cm, clip=true, width=0.9\linewidth]{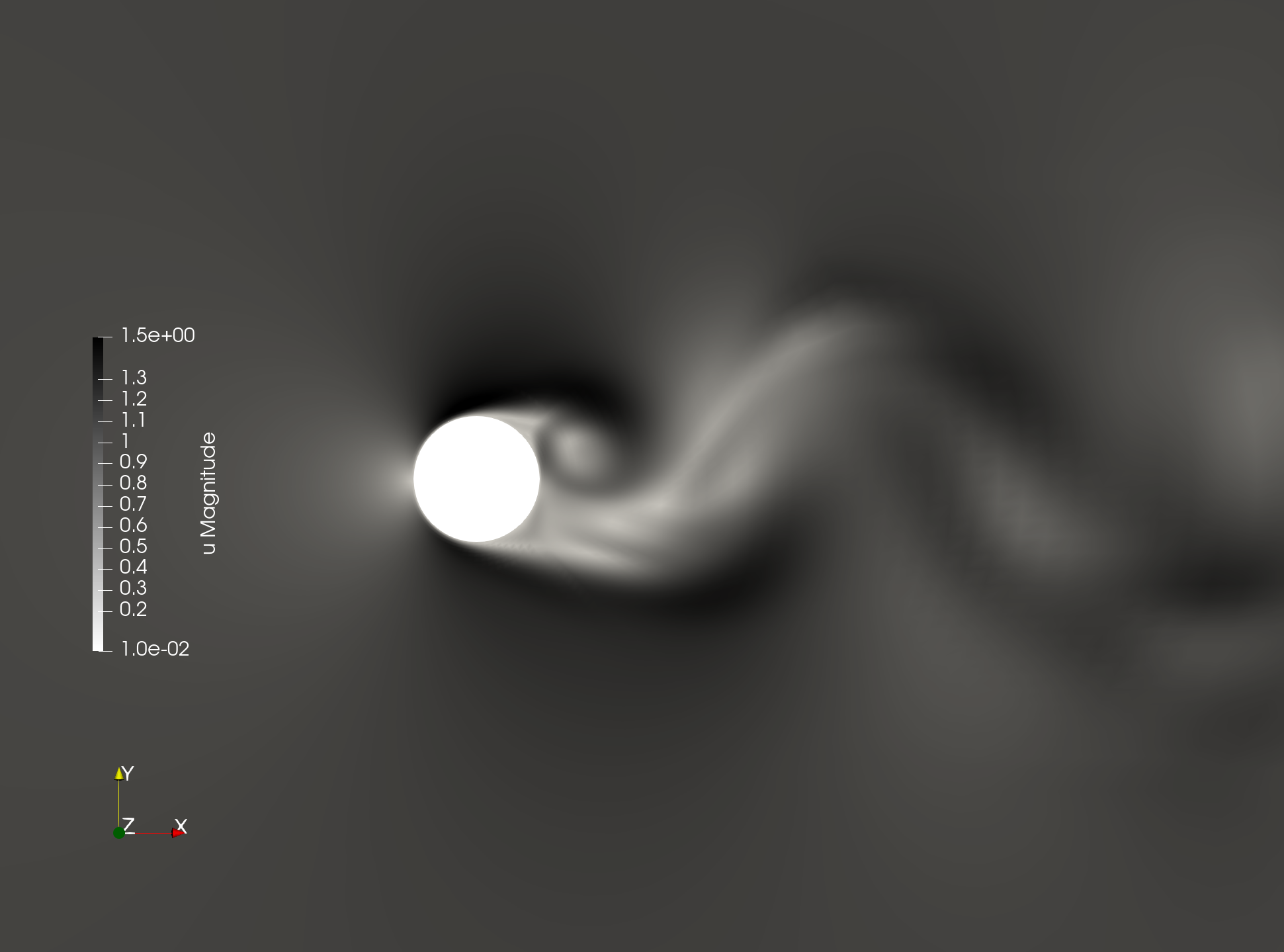}
    \includegraphics[width=0.9\linewidth]{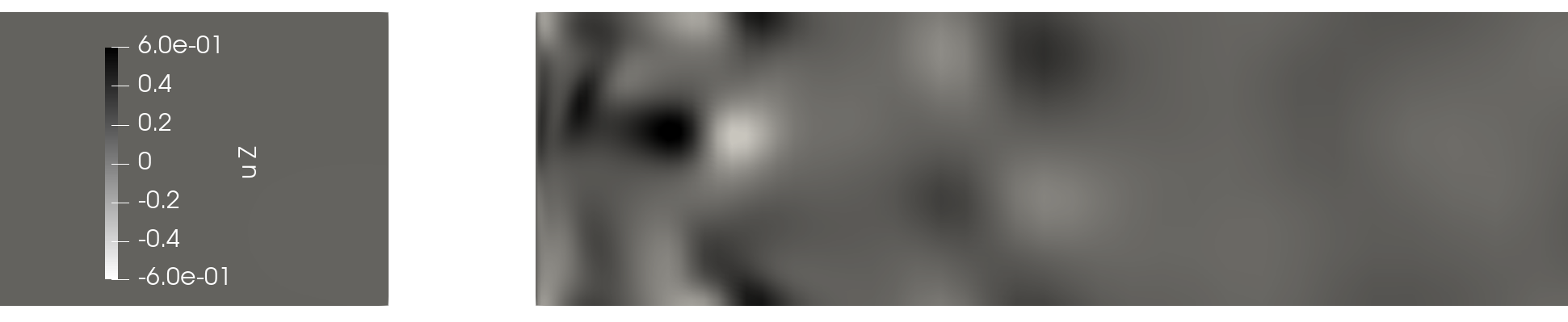}
    \caption{
       Instantaneous visualization of the flow field. 
       Top: vertical cross-section, plotted by the magnitude of velocity. 
       Bottom: horizontal cross-section, plotted by the span-wise velocity. 
       The bottom picture shows the flow is 3D. 
       All velocities are normalized by the reference velocity $u_r$.
  }
    \label{f:cylinder_velocity}
\end{figure}

% A snapshot of primal flow fields
A snapshot of the flow field simulated with above settings is shown in figure~\ref{f:cylinder_velocity}.
As we can see, this flow exhibits weak turbulence in the wake.
In particular, the top view shows that this flow is 3D.

% parameter, objective 
We choose our system parameter as the Mach number of the incoming flow.
The objective function is the time-averaged normalized drag over the cylinder. 
More specifically,
\begin{equation}
  J_{avg} = \frac{2}{\rho_r u_r^2 zd}\avgT \int (pn_x -\mu\bm{n}\cdot\nabla u_x) \,dS\,dt,.
\end{equation}
Here the second integral is over the surface of the cylinder;
$u_r = 31.4 m/s$ and $\rho_r = 1.3 kg/m^3$ are the reference velocity and density of the base case, where $Ma=0.093$.
For the base case, the normalized drag and the drag coefficient are the same, whereas for other Mach numbers they are different.

\begin{figure}[htb!]
  \centering
  \includegraphics[width=0.7\linewidth]{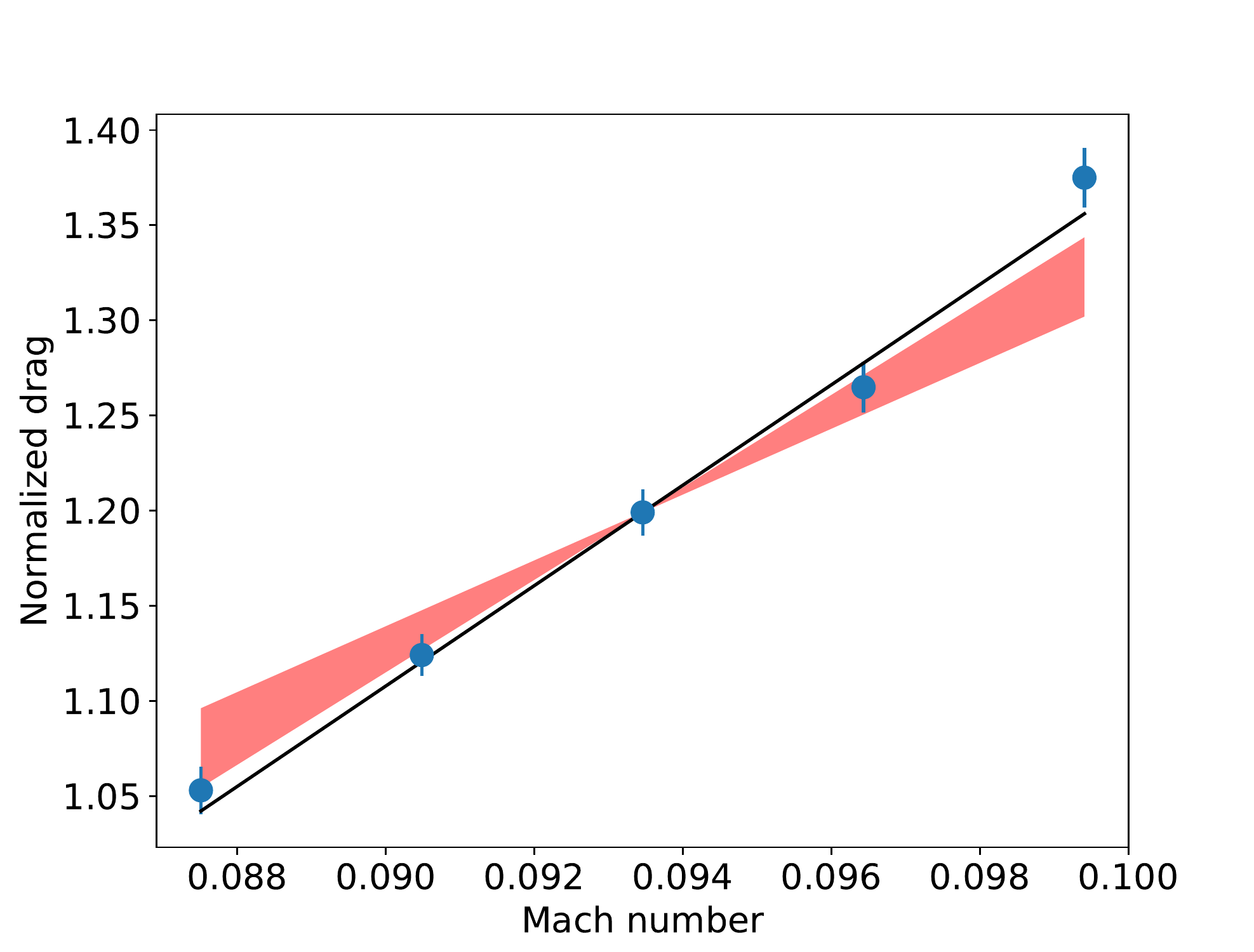} 
  \caption{Normalized drag as a function of inlet Mach number. 
    Blue bars denote the confidence interval of the averaged normalized drag.
    The black line denotes the sensitivity estimated using linear regression. 
    The red shaded region denotes the confidence interval of the sensitivity estimated using NILSAS. }
  \label{f:cylinder_error1}
\end{figure}

% show objectives we obtain: 1) base case 2) different Ma
We run the flow simulation for $10^6$ time steps, which corresponds to approximately $720 t_r$.
Here the time unit $t_r$ is the amount of time that the flow takes to traverse the length of the cylinder, that is, $t_r = d/u_r$. 
This time interval is sufficient to obtain a statistically converged estimate of the design objective.
The standard deviation of the time-averaged objective is computed using
the autoregressive time series analysis techniques described in \cite{talnikar2015optimization} and \cite{oliver2014},
and we use one standard deviation as the confidence interval.
The normalized drag for the base case is $1.2 \pm 0.03$.
Our results reasonably matches the results from experiments \cite{wieselsberger1922new,roshko1961experiments,tritton1959experiments},
which is approximately $1.0 \pm 0.15$.
Figure \ref{f:cylinder_error1} shows different objectives for different incoming Mach number.

% sensitivity using regression.
We first estimate the sensitivity by the linear least-squares regression method using $5$ data points with different parameter values.
To use this linear regression method, we need to make the assumption that the relation between parameters and objectives is linear.
We select one standard deviation of the relevant estimator as our confidence interval.
Note that one shortcoming of the linear regression method is that the linear assumption may not be true when parameters are spaced far apart
such that a linear approximation no longer holds on the dataset;
on the other hand, when parameters are too close, the uncertainty in the objectives will lead to large error in the sensitivity.
In the base case, the sensitivity of drag with respect to the inflow Mach number, given by linear regression, is $25.0 \pm 2.1$. 
This sensitivity is visualized in figure~\ref{f:cylinder_error1}.

\begin{figure}[htb!]
    \centering
    \includegraphics[width=0.75\linewidth]{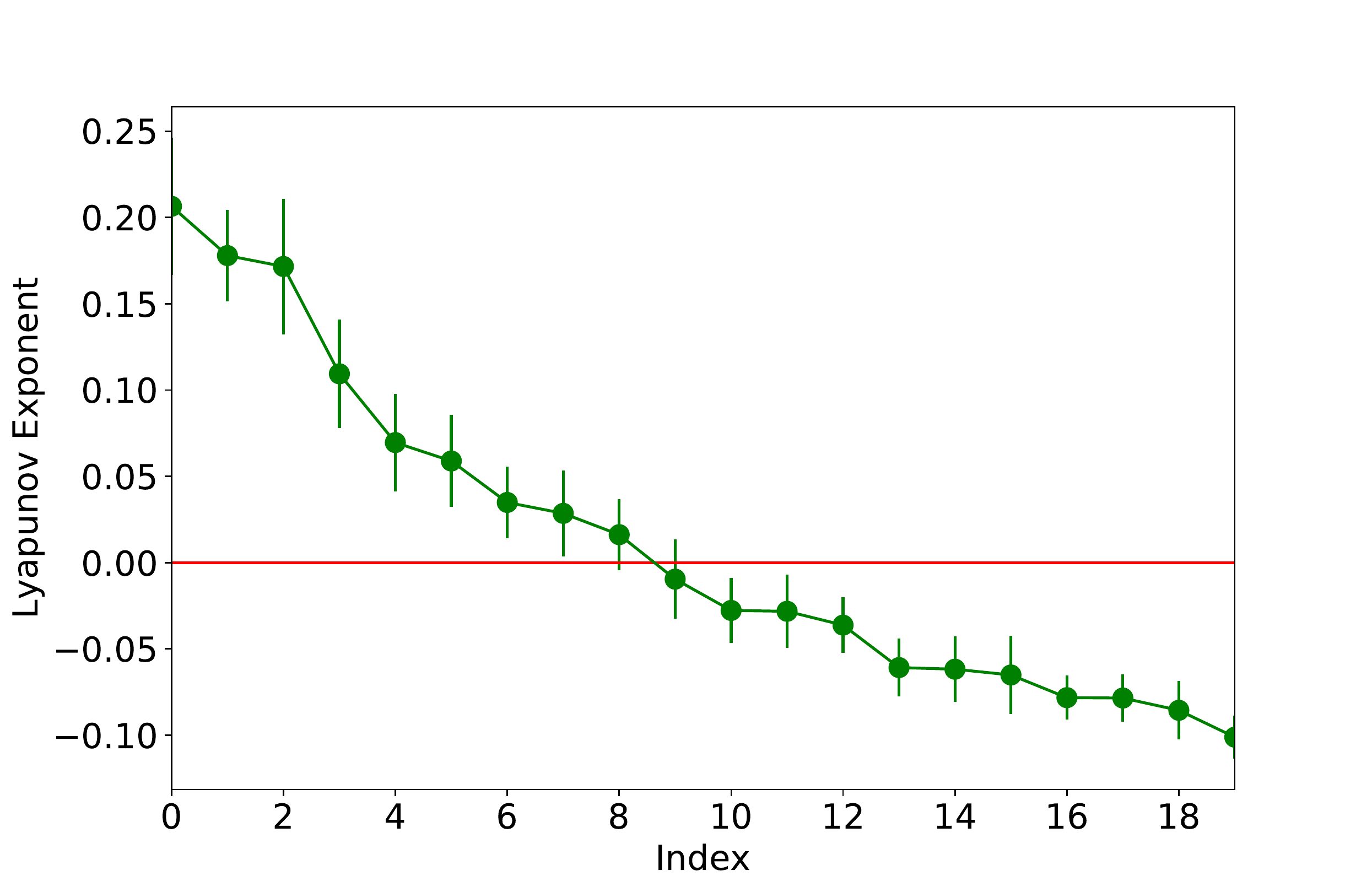}
    \caption{
      Spectrum of the first 20 adjoint Lyapunov Exponents (LE). The time unit for LEs is $t_r^{-1}$.  
      The largest LE is $0.21t_r^{-1}$, meaning in one time unit $t_r$, the norm of the first adjoint CLV becomes $e^{0.21} = 1.23$ times larger.
    }
    \label{f:lyapunov_exponents}
\end{figure}

% NILSAS parameters: M
% other NILSAS parameters
Adjoint LEs are shown in figure~\ref{f:lyapunov_exponents},
where the confidence interval is also selected as one standard deviation given by autoregressive time series analysis.
The subsonic flow over a 3D cylinder has $m_{us} = 9$ unstable adjoint CLVs.
%Our adjoint LE spectrum of the cylinder flow at $Re=1100$ is comparable to 
%the tangent LE spectrum of the cylinder at $Re=525$ in \cite{Ni_CLV_cylinder}:
%for example, the largest adjoint LE in our paper is $0.21t_r^{-1}$, whereas the largest tangent LE in \cite{Ni_CLV_cylinder} is $0.20t_r^{-1}$. 
%Although the two cases are different in Reynolds numbers and numerical solvers,
%still we can tell from numerical results that the tangent and adjoint LE spectrum should have similarity.
In NILSAS, the number of homogeneous adjoint solutions computed is set to $M = 20$.
The number of segments in NILSAS is $K=100$ and the number of time steps per segment is $500$. 
Each segment roughly corresponds to $0.4t_r$. 
Consequently, the time length of trajectory used in NILSAS is $40t_r$, 
which is much lower than that required to obtain a reasonably accurate sensitivity using the linear regression method.
In this particular implementation, corresponding to the discussion in section \ref{s:remarks misc},
on segment $[t_i, t_{i+1}]$, we approximate integrations in equation~\eqref{e:covariant matrix} using snapshots at $t_{i+1}$.
As a result, we have $C_i=I$, $d_i^{wv^*} = 0$:
this approximation eases the implementation responsible for storing adjoint solutions.

% NILSAS results
The sensitivity computed by NILSAS is $20.8 \pm 3.5$, which is visualized in figure \ref{f:cylinder_error1}.
Here the confidence interval is also selected as one standard deviation given by autoregressive time series analysis.
Comparing to the sensitivity estimated via linear regression methods, the relative difference is less than $20\%$. 
As we can see, the sensitivity computed by NILSAS correctly reflects the trend between parameters and objectives.

% this system is not very good hyperbolic
We remark that NILSAS may work for systems do not strictly satisfy the assumptions in theorem~\ref{thm:flow},
and our fluid problem is such an example.
First, our system has at least two neutral CLVs:
the first one corresponds to the common time translation of continuous dynamical systems,
and the second corresponds to span-wise translations due to the periodic boundary conditions.
Second, due to the similarity of this fluid problem with the one investigated in \cite{Ni_CLV_cylinder},
whose tangent CLVs appear to have occasional tangencies,
it is reasonable to assume that adjoint CLVs in our current system may also have occasional tangencies.
Still, like NILSS did in \cite{Ni_CLV_cylinder},
NILSAS computes a correct sensitivity: this encourages us to test NILSS and NILSAS on more general chaotic systems.

% Cost comparison and analysis. Talk about scaling to larger turbulent flow problems
We compare computational cost for sensitivity analysis via the linear regression method and NILSAS.
The linear regression method runs the primal solver for a total of $5\times 10 ^6$ steps.
NILSAS runs the primal solver and 21 adjoint solvers, each for $5\times 10 ^4$ steps, which leads to
 $1.1\times 10^6$ steps in total.
To build more favor towards NILSAS, note that,
first, adjoint solvers can be further accelerated due to the vectorization we discussed in section~\ref{s:remarks misc};
second, NILSAS has no additional cost for sensitivities to multiple parameters.
For chaotic problems with a higher number of positive LEs, 
the cost of NILSAS increases;
however, if the percentage of positive LEs is still low,
the non-intrusive formulation can still be a key technique for designing fast sensitivity algorithms.

\section{Conclusions}

% how we developed NILSAS.
To compute the gradient of long-time averaged objectives in chaotic systems,
we develop the Non-Intrusive Least Squares Adjoint Shadowing (NILSAS) algorithm, 
which approximates the adjoint shadowing direction by a `non-intrusive' formulation of a least squares problem.
NILSAS is demonstrated on the Lorenz 63 system and a weakly turbulent 3D flow over a cylinder,
where it gives accurate sensitivities for both cases.

% compare to NILSS and adjoint NILSS
Similar to NILSS \cite{Ni_NILSS_JCP}, NILSAS can be implemented with little modification to existing adjoint solvers,
and its minimization is carried out only in the unstable adjoint subspace.
Unlike NILSS, NILSAS has the benefit of adjoint approaches that its cost does not increase with the number of parameters;
thus making NILSAS ideal for applications where there are many parameters, or where $f_s$ is unknown a priori.
NILSAS does not require tangent solvers, and is easy to implement.

\appendix

\section{Solving the NILSAS problem}\label{app:solve_NILSAS}

We discuss one way to solve the NILSAS problem in equation~\eqref{e:NILSAS_on_multiple_segments}.
The corresponding Lagrange function is:
\begin{equation}\begin{split}
  &\sum_{i=0}^{K-1} \frac 12 (a_i)^T C_i a_i + (d^{wv^*}_i)^T a_i  \\
  + &\sum_{i=1}^{K-1} \lambda_i^T \left( a_{i-1} - R_i a_i - b_i \right)
  + \lambda' \left( \sum_{i=0}^{K-1} (d^{wf}_i)^T a_i + \sum_{i=0}^{K-1} d^{v^*f}_i \right) \,,
\end{split}\end{equation}
where $\lambda_i$ is the Lagrange multiplier for the continuity condition at $t_i$.
By the Lagrange multiplier method,
the minimizer for the NILSAS problem is at the solution of the following linear equation systems:
\begin{equation}\label{e:NILSAS_block_matrix_form}
  \mat{ C & B^T\\ B & 0 }  
  \mat{ a\\  \lambda } 
  =\mat{ -d \\ b }\;,
\end{equation}
where the block matrices $C\in\R^{MK \times MK}$, $B\in \R^{(MK-M+1)\times MK}$,
vectors $a, d\in \R^{MK}$ and $\lambda, b \in \R^{MK-M+1}$.
More specifically, 
\begin{equation}\begin{split}
  & C = \mat{C_0 \\ &C_1 \\&& \ddots \\ &&&C_{K-1}}
  , \quad
  B = \mat{I &-R_1\\ &I &-R_2 \\&&\ddots &\ddots \\ &&& I &-R_{K-1}\\
  (d^{wf}_0)^T &&\cdots & &(d^{wf}_{K-1})^T}, \\
  & a = \mat{a_0 \\ \vdots\\a_{K-1}}
  , \quad
  \lambda = \mat{\lambda_1 \\ \vdots\\ \lambda_{K-1} \\ \lambda'}
  , \quad
  d = \mat{d^{wv^*}_0 \\ \vdots\\d^{wv^*}_{K-1}}
  , \quad
  b = \mat{b_1 \\ \vdots\\b_{K-1} \\ -  \sum_{i=0}^{K-1} d^{v^*f}_i}
  ,
\end{split}\end{equation} 
where $\{C_i\}_{i=0}^{K-1}$, $\{R_i\}_{i=1}^{K-1} \subset \R^{M\times M}$;
$\{a_i\}_{i=0}^{K-1}$, $\{d^{wf}_i\}_{i=0}^{K-1}$, $\{d^{wv^*}_i\}_{i=0}^{K-1}$, $\{\lambda_i\}_{i=1}^{K-1}$, $\{b_i\}_{i=1}^{K-1}\subset \R^M$;
$\lambda'$, $ \{d^{v^*f}_i\}_{i=0}^{K-1} \subset \R$.

We can solve the Schur complement of equation~\eqref{e:NILSAS_block_matrix_form} for $\lambda$:
\begin{equation}
  - B C^{-1} B^T \lambda = B C^{-1}d + b \;,
\end{equation}
where $C^{-1}$ can be computed via inverting each diagonal block in $C$.
Then we can compute $a$ by:
\begin{equation}
  a = - C^{-1}(B^T \lambda +d) \;.
\end{equation}

\section{NILSAS on discrete systems}\label{s:discrete systems}

\subsection{Backgrounds and notations}

We provide a brief introduction on discrete dynamical systems, in particular, hyperbolic diffeomorphisms.
More details are provided in \cite{Ni_adjoint_shadowing}.
First, the governing equation for a discrete dynamical system is: 
\begin{equation} \label{e:primal_system_diffeo}
  u_{l+1} = f(u_l,s), \quad l\ge 0\,.
\end{equation}
The objective is:
\begin{equation} \label{eq:average_J_diffeo}
  \avg{J}:= \lim\limits_{N\rightarrow\infty} \frac{1}{N} \sum_{l=0}^{N-1} J(u_l,s).
\end{equation}
Similar to flows, we assume $u_l\in \R^m$, and $f(u,s)$ and $J(u,s)$ are smooth.
We also assume $f$ is a diffeomorphism in $u$, that is, for each fixed $s$, $f$ has a smooth inverse.
Also, for simplicity of notations, we assume there is only one parameter $s\in\R$.

We first look at the tangent equations.
The homogeneous tangent diffeomorphism is: 
\begin{equation}\label{e:homo_tangent_diffeo}
  w_{l+1} = f_{ul} w_l \,.
\end{equation}
where the second subscript of $f_{ul}$ indicate where the partial derivative is evaluated, that is,
$f_{ul}:=$ $\partial f/ \partial u (u_l,s)$.
A tangent CLV with exponent $\lambda$ is a homogeneous tangent solution $\{\zeta_l\}^\infty_{i=0}$ such that there is constant $C$,
for any integer $l_1, l_2$, $\| \zeta_{l_2} \| \le C e^{\lambda(l_2-l_1)}\| \zeta_{l_1} \|$. 
The uniform hyperbolicity for diffeomorphisms is defined as that all LEs are not 1.

% main definitions for adjoint flows
On the adjoint side, the homogeneous adjoint diffeomorphism is defined as:
\begin{equation} \label{e:homo_adjoint_diffeo}
  \aw_{l} = f_{ul}^T \aw_{l+1} \,,
\end{equation}
where $\cdot^T$ is the matrix transpose.
The particular inhomogeneous adjoint diffeomorphism we will be using is:
\begin{equation} \label{e:inhomo_adjoint_diffeo}
  \av_{l} = f_{ul}^T \av_{l+1} + J_{ul} .
\end{equation}
On a trajectory $\{u_l\}_{l=0}^{\infty}$ on the attractor,
the adjoint shadowing direction $\{\av_l\}_{l=0}^{\infty}$ is a sequence with the following properties:
\begin{enumerate}
  \item $\{\av_l\}_{l=0}^{\infty}$ solves an inhomogeneous adjoint equation:
    \begin{equation} 
      \av_{l} = f_{ul}^T \av_{l+1} + J_{ul} \,, 
    \end{equation}
  \item $\av_0$ has zero component in the unstable adjoint subspace. 
  \item $\|\av_l\|$ is bounded by a constant for all $l\ge 0$.
\end{enumerate}

% main theorem
It was proved in \cite{Ni_adjoint_shadowing} that for a uniform hyperbolic diffeomorphism with a global compact attractor,
on a trajectory on the attractor, there exists a unique adjoint shadowing direction.
Further, we have the adjoint sensitivity formula:
\begin{equation}\label{e:djds_adjoint_diffeo}
  \dd {\avg{J}} {s} = \lim_{N\rightarrow \infty} \frac 1N \sum_{l=0}^{N-1} \left( \ip{\av_{l+1}, f_{sl}} + J_{sl} \right) \,.
\end{equation}

\subsection{Procedure list of NILSAS}

% what solver we need
We provide a procedure list for the NILSAS algorithm on discrete chaotic systems, more specifically, hyperbolic diffeomorphisms.
To start with, we need an inhomogeneous adjoint solver and a homogeneous adjoint solver, both can take arbitrary terminal conditions.
The inhomogeneous adjoint equation still has right-hand-side $-J_u$, same as many existing adjoint solvers for discrete systems.
We provide the following data: 
1) the number of homogeneous adjoint solutions, $M\ge\mus$, where $\mus$ is the number of unstable CLVs, 
note that because the lack of neutral CLV, we can use one less homogeneous adjoint solution; 
2) the total number of segments, $K$; 
3) number of steps in one segment, $L$.

% subscripts explanation
We can have three subscripts, the first, typically being $u$ or $s$, indicates this term is a partial derivative;
the second, typically being $i$, $0$, or $K$, indicates the segment number;
the third, typically being $l$, $0$, or $L$, indicates the step number inside a segment.
Disappearance of the first subscript means that term is not a partial derivative.
Disappearance of the third subscript means either we are considering all steps in a segment, 
or that term is defined only once per segment interface.

\begin{enumerate}

  \item Integrate the primal system for sufficiently many steps so that the initial condition, $u_{00}$, is on the attractor.

  \item Compute the trajectory $u_{il}$ for $0\le i\le K-1$ and $0\le l \le L$.
    Here we assume the step at end of each segment overlaps with the start of next segment,
    that is, $u_{iL} = u_{i+1,0}$.

  \item Generate terminal conditions for $\aW_{i}$ and $\av^*_i$ on the last segment $i=K-1$:
  \begin{enumerate}
    \item Randomly generate a $m\times M$ full rank matrix, $Q'$.
      Perform QR factorization: $Q_K R_K = Q'$.
    \item Set $p_{K} = 0$.
  \end{enumerate}

  \item Compute $\aW_i$ and $v^*_i$ on all segments.
  For $i=K-1$ to $i=0$ do:
  \begin{enumerate}
    \item 
      To get $\aW_{il}$, whose columns are homogeneous adjoint solutions on segment $i$, solve:
      \begin{equation}
        \aw_{il} = f_{uil}^T \aw_{i,l+1} \,, \quad \aW_{iL} = Q_{i+1} \,.
      \end{equation}
      To get $\av^*_i(t)$, solve the inhomogeneous adjoint equation:
      \begin{equation}
        \av_{il} = f_{uil}^T \av_{i,l+1} + J_{uil} \,,\quad \av^*_{iL} = p_{i+1} \,.
      \end{equation}

    \item Compute the following integrations.
      \begin{equation} \begin{split}
        &C_i = \sum_{l=1}^{L} \aW_{il}^T \aW_{il} dt \,, \quad
        d^{wv^*}_i = \sum_{l=1}^{L} \aW_{il}^T \av^*_{il} dt \,, \quad
        d^{J_s}_i = \sum_{l=1}^{L} J_{sil} dt \,,\\
        &d^{wf_s}_i =\sum_{l=1}^{L} \aW_{il}^T f_{si,l-1} dt \,, \quad
        d^{v^*f_s}_i = \sum_{l=1}^{L} \av^{*T}_{il} f_{si,l-1} dt \,,
      \end{split} \end{equation}
      where $d^{wv^*}_i$, $d^{wf_s}_i\in \R^M$; $d^{v^*f_s}_i$, $d^{J_s}_i\in\R$;
      $C_i\in \R^{M\times M}$ is the covariant matrix.
      Note that when multiplying adjoint solutions with $f_s$, their time steps are not the same:
      this asymmetry is the same as that in equation~\eqref{e:djds_adjoint_diffeo}.
      We are not sure yet if this technical detail can be neglected in practice.
    
    \item Orthonormalize homogeneous adjoint solutions via QR factorization:
      \begin{equation}
        Q_i R_i = \aW_{i0}
      \end{equation}

    \item Rescale the inhomogeneous adjoint solution using $Q_i$: 
      \begin{equation}
        p_i = \av^*_{i0} - Q_i b_i \,,\quad
        \textnormal{where } b_i = Q_i^T \av^{*}_{i0} \,.
      \end{equation}
  \end{enumerate}

\item Compute the adjoint shadowing direction $\{\av_{il}\}$ for  $0\le i\le K-1$ and $0\le l \le L$.
  \begin{enumerate}
    \item Solve the NILSAS problem on multiple segments:
      \begin{equation} 	\begin{split}
        \min_{a_0,\cdots,a_{K-1} \in \R^M} &\sum_{i=0}^{K-1} \frac 12 (a_i)^T C_i a_i +  (d^{wv^*}_i)^T a_i, \quad  \mbox{s.t.}\\
        &a_{i-1} = R_i a_i + b_i \,, \quad  i=1,\cdots,K-1 \;.
      \end{split}\end{equation}
      This is a least squares problem in $\{a_i\}_{i=0}^{K-1} \subset \R^M$.
      Note we do not have the other constraint as NILSAS in the continuous case.

    \item On each time segment $i$, $\av_{il} $ is given by
      \begin{equation}
        \av_{il} = \av^*_{il} + \aW_{il} a_i .
      \end{equation}
  \end{enumerate}

  \item Compute the derivative by:
    \begin{equation} 
      \frac{d\avg{J}}{ds} \approx
      \frac 1{KL} \sum_{i=0}^{K-1} \left( d^{v^*f_s}_i + a_i^T d^{wf_s}_i + d^{J_s}_i \right)
    \end{equation}
\end{enumerate}

\bibliographystyle{model1-num-names}
\bibliography{MyCollection}

\end{document}